\documentclass[tighten,twocolumn]{aastex61}

\usepackage{longtable}
\usepackage{afterpage}

\usepackage{graphicx}
\usepackage{rotating}
\usepackage{float}

\usepackage{epstopdf}

\begin{document}

\shorttitle{Magnetic interactions in the solar atmosphere observed by \textit{IRIS}. I.}
\shortauthors{Guglielmino et al.}

\title{\textit{IRIS} observations of magnetic interactions in the solar atmosphere\\between pre-existing and emerging magnetic fields. I. Overall evolution.}

\author{Salvo L. Guglielmino}
\affiliation{Dipartimento di Fisica e Astronomia - Sezione Astrofisica, Universit\`{a} degli Studi di Catania,
	Via S.~Sofia 78, 95123 Catania, Italy}

\author{Francesca Zuccarello}
\affiliation{Dipartimento di Fisica e Astronomia - Sezione Astrofisica, Universit\`{a} degli Studi di Catania,
	Via S.~Sofia 78, 95123 Catania, Italy}

\author{Peter R. Young}
\affiliation{College of Science, George Mason University, Fairfax, VA 22030, USA}
\affiliation{Code 671, NASA Goddard Space Flight Center, Greenbelt, MD 20771, USA}
\affiliation{Northumbria University, Newcastle upon Tyne, NE1 8ST, UK}

\author{Mariarita Murabito}
\affiliation{Dipartimento di Fisica e Astronomia - Sezione Astrofisica, Universit\`{a} degli Studi di Catania,
	Via S.~Sofia 78, 95123 Catania, Italy}

\author{Paolo Romano}
\affiliation{INAF - Osservatorio Astrofisico di Catania,
	Via S.~Sofia 78, 95123 Catania, Italy}

\correspondingauthor{Salvo L. Guglielmino}
\email{salvo.guglielmino@oact.inaf.it}

\begin{abstract}

We report multi-wavelength ultraviolet observations taken with the \textit{IRIS} satellite, concerning the emergence phase in the upper chromosphere and transition region of an emerging flux region (EFR) embedded in the pre-existing field of active region NOAA 12529. \textit{IRIS} data are complemented by full-disk observations of the \textit{Solar Dynamics Observatory} satellite, relevant to the photosphere and the corona. The photospheric configuration of the EFR is also analyzed by measurements taken with the spectropolarimeter aboard the \textit{Hinode} satellite, when the EFR was fully developed. Recurrent intense brightenings that resemble UV bursts, with counterparts in all coronal passbands, are identified at the edges of the EFR. Jet activity is also observed at chromospheric and coronal levels, near the observed brightenings. The analysis of the \textit{IRIS} line profiles reveals the heating of dense plasma in the low solar atmosphere and the driving of bi-directional high-velocity flows with speed up to $100 \,\mathrm{km\,s}^{-1}$ at the same locations. Compared with previous observations and numerical models, these signatures suggest evidence of several long-lasting, small-scale magnetic reconnection episodes between the emerging bipole and the ambient field. This process leads to the cancellation of a pre-existing photospheric flux concentration and appears to occur higher in the atmosphere than usually found in UV bursts, explaining the observed coronal counterparts.

\end{abstract}


\section{Introduction}

During the last few years, data acquired by the \textit{Interface Region Imaging Spectrograph} \citep[\textit{IRIS},][]{DePontieu:14} have provided a new insight into several phenomena that occur in the solar atmosphere, specially in the upper chromosphere and in the transition region (TR). The latter is the atmospheric layer characterized by an abrupt increase of temperature, from chromospheric values ($\approx 10^4\,\mathrm{K}$) up to $10^5\,\mathrm{K}$ over a distance of some tens of km, reaching the typical coronal values of $1 - 2$~MK in about 2000~km. 

Investigations of small-scale energy release episodes, such as explosive events \citep[EE,][]{Huang:14,Gupta:15,Huang:17}, penumbral jets and brightenings \citep{Tian:14,Vissers:15b,Bai:16,Deng:16,Alissandrakis:17,Samanta:17}, and EUV jets \citep{Chen:16}, have largely benefitted from \textit{IRIS} observations. In addition, UV bursts have been discovered \citep{Peter:14}. These transient events, also called \textit{IRIS} bombs, show in UV lines a several-order of magnitude increase in radiation and plasma flows of hundreds of $\mathrm{km\,s}^{-1}$, occurring on spatial scales of $\approx 500$~km for a short time ($\sim 5$ minutes). They are thought to be caused by small-scale magnetic reconnection episodes that heat plasma up to $\approx 10^5\,\mathrm{K}$, driving bidirectional high-speed plasma flows. Interestingly, these events seem to occur at low atmospheric heights \citep{Peter:14,Grubecka:16}. That has led to debate on their relationship with Ellerman bombs \citep[EBs; see][]{Kim:15,Vissers:15a,Tian:16,Hong:17,Zhao:17}. 

The presence at the photospheric level of opposite magnetic polarities that come into contact and/or cancel with each other appears to be the common denominator in all of those phenomena that are characterized by small-scale transient brightenings and jet-like ejections \citep{Shimizu:15}. In particular, reconnection may occur when an emerging flux region (EFR) interacts with the overlying, pre-existing field lines, thus triggering high-temperature emission in localized regions and surge/jet ejections \citep[see, e.g.,][]{Guglielmino:12,Cheung:14}. Indeed, this scenario has been reported in very detailed high-resolution observations of small-scale EFRs and of their chromospheric and coronal response, carried out in recent years \citep[e.g.][]{Guglielmino:08,Guglielmino:10,Santiago:12,Ortiz:14,Jaime:15,Centeno:17}. In this respect, it is necessary to carry out the investigation of the evolution of EFRs and of their interaction with the surrounding atmosphere at several wavelengths, in order to obtain a complete picture of the phenomena taking place at these locations.

In connection with these observations, numerical radiative magnetohydrodynamic (MHD) simulations of magnetic flux emergence in the solar atmosphere show that small-scale energy release events occur ubiquitously in the EFRs, as a result of magnetic reconnection. This is found at EFR locations either owing to the interaction of the newly emerged magnetic flux with the pre-existing ambient field \citep{Shibata:89,Archontis:04,Archontis:05,Archontis:07,Galsgaard:05,Galsgaard:07,Isobe:07,Isobe:08,Sykora:08,Sykora:09,Archontis:09,Tortosa:09,Cheung:10,Hood:12,David:15,Syntelis:15,Ni:15,Ni:16,Nobrega:16,Nobrega:17} or to the self-interaction of the emerging bipolar flux concentrations \citep[e.g.,][]{Hansteen:17}. These models demonstrate that the reconnection process can supply enough energy to heat and accelerate the plasma. They also illustrate that the dynamics and energetics of the process is complex in three-dimensional geometry, due to the relative orientation between the emerging field lines and the pre-existing field \citep[e.g.,][]{Galsgaard:07}. Joule heating seems to be the predominant mechanism to energize the plasma in regions with weak magnetic fields and high plasma $\beta$, whereas slow-mode and fast-mode shocks appear to be the main mechanisms in a strong magnetic environment, with low plasma $\beta$ \citep{Ni:16}.

The advent of \textit{IRIS} has further extended our capabilities to scan the various layers of the Sun with simultaneous multi-wavelength observations during flux emergence episodes. In fact, now we can study the consequences of magnetic interactions between the newly emerged flux and the ambient field in the TR as well \citep{Santiago:14,Jiang:15,Ortiz:16}. In addition to the detection of brightness enhancements and plasma ejections in the vicinity of EFRs, this also includes the possibility to detail the connectivities of the emerging magnetic field in the solar chromosphere and TR, and deriving some signatures of the rising plasma, e.g., temporal delays between the passage through different atmospheric heights and vertical velocities through imaging and spectroscopic information, respectively. In this perspective, a recent study by \citet{Toriumi:17} has reported on heating events observed during the earliest phase of flux emergence of an EFR in active region (AR) NOAA 12401, analyzing simultaneous observations by \textit{IRIS}, \textit{Hinode}, and the \textit{Solar Dynamics Observatory} \citep[SDO,][]{Pesnell:12} satellites. \citet{Toriumi:17} have found that some \ion{Ca}{2} H bright points in the EFR center, cospatial with regions of mixed polarities in the photosphere, have \textit{IRIS} UV spectra that exhibit flare-like light curves and enhanced red- or blue-shifted tails with velocity up to $\pm 150 \,\mathrm{km \,s}^{-1}$, suggesting the presence of bi-directional jets. 

In spite of the advances in our knowledge of the physical processes responsible for the interaction of EFRs with the ambient magnetic field obtained during the last decades thanks to multi-wavelength high-resolution observations, some issues remain open. It is not yet understood why episodic brightenings and jets are not always observed in the emergence sites: is this due to the respective orientation of the interacting magnetic fields, to the magnetic energy carried by the emerging flux, or to an already unstable configuration of the pre-existing magnetic field? Here, we present the analysis of the evolution of an EFR embedded in a unipolar plage in the photosphere and the response of the overlying atmospheric layers to the emerging flux, in particular of the upper chromosphere and TR, using high-resolution observations by the \textit{IRIS} and \textit{Hinode} satellites. This study is complemented with data from the \textit{SDO} mission, which reveal coronal counterparts of the phenomena occurring during the emergence of the EFR. 

In the next Section we describe the observations and the data analysis. In Sect.~3 we present our findings, which are discussed in Sect.~4. Section~5 summarizes our conclusions in a more general context.


\section{Observations and data analysis}

\begin{figure}[t]
	\centering
	\includegraphics[scale=0.4, clip, trim=55 20 190 10]{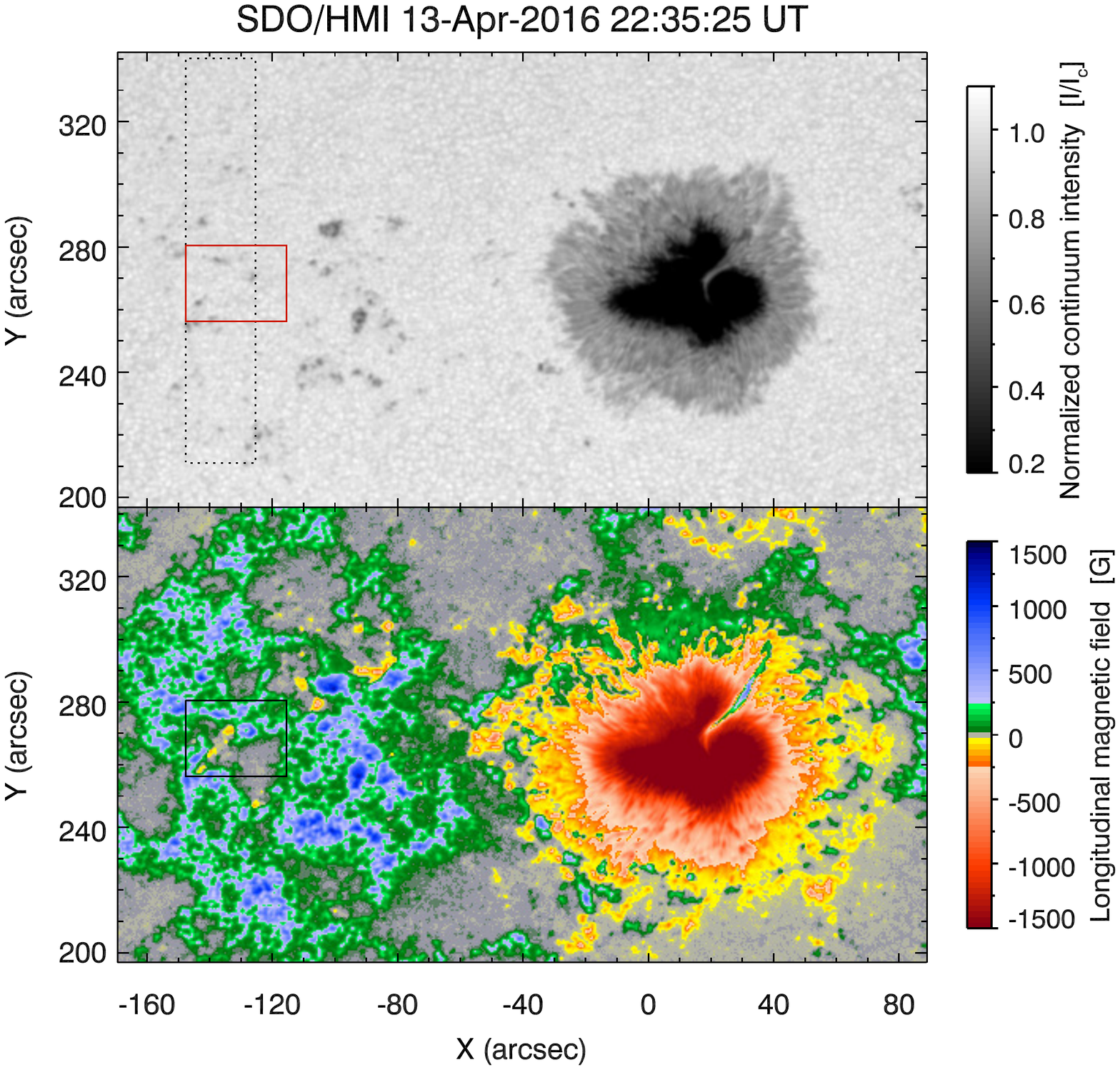}
	\caption{AR NOAA 12529 as seen in the SDO/HMI continuum filtergram and in the simultaneous map of the LOS component of the magnetic field, closest in time to the beginning of \textit{IRIS} observations at 22:34:43~UT. The solid box frames the portion of the FoV where the EFR appears. The dashed box indicates the area covered by the \textit{IRIS} slit during the six large dense 64-step rasters. Here and in the following figures, North is at the top, West is to the right. The axes give the distance from solar disc center. \label{fig_context}}
\end{figure} 

During April 2016, the prominent AR NOAA 12529 appeared on the Sun's visible hemisphere, characterized by a $\beta$-type magnetic configuration \citep{Guglielmino:17}. Between April 13 and 14 it passed across the central meridian, being located at heliocentric angle $\mu \approx 0.96$. At that time, an EFR was emerging within the following positive polarity of the AR.

With regard to the observations of the EFR at the photospheric level, we analyzed continuum filtergrams and line-of-sight (LOS) magnetograms taken by the Helioseismic and Magnetic Imager \citep[HMI,][]{Scherrer:12} on board the \textit{SDO} satellite along the \ion{Fe}{1}~6173~\AA{} line, with a spatial resolution of 1\arcsec. These full-disk data cover about 12 hours of observations, starting from April~13 at 19:12~UT until April~14 at 07:00~UT, with a cadence of 45~s. The field-of-view (FoV) used for our analysis is shown in Figure~\ref{fig_context} for the continuum and longitudinal component of the magnetic field. In the maps, a solid box frames the area occupied by the EFR.

At a later time during the evolution of the EFR, we also benefitted from photospheric observations performed by the spectropolarimeter \citep[SP,][]{Lites:13} of Solar Optical Telescope \citep[SOT,][]{Tsuneta:08} aboard the \textit{Hinode} \citep{Kosugi:07} satellite. This instrument obtained a single raster scan of AR NOAA 12529 from 02:21 to 03:24~UT on April~14 along the \ion{Fe}{1} line pair at 6302~\AA{}. This SOT/SP scan has a pixel size along the slit of 0\farcs32, with a step size of 0\farcs32, and a step cadence of 3.8~s (Fast Mode). The region was scanned in 1000 steps, covering a FoV of about $300 \arcsec \times 162\farcs3 $. 

To remove the stray light contamination induced by the spatial point spread function (PSF) of the telescope, we performed a deconvolution of the original data using a regularization method based on a principal component decomposition of the Stokes profiles, as proposed by \citet{PCA:13}. We followed the implementation of this method for \textit{Hinode} SOT/SP data by \citet{Quintero:16}. The photospheric vector magnetic fields were obtained by applying the SIR inversion code \citep{SIR:92} to the deconvolved SOT/SP data. We used the Harvard Smithsonian Reference Atmosphere \citep[HSRA;][]{HSRA} as the initial model. The inversion yielded the temperature stratification in the range $-4.0 < \log \tau_{5000} < 0$ ($\tau_{5000}$ is the optical depth of the continuum at 5000~\AA). SIR also provided the LOS velocity, the micro-turbulent velocity, the magnetic field strength $B$, as well as the inclination and azimuth angles $\gamma$ and $\phi$ in the LOS reference frame. These quantities were assumed to be constant with optical depth. As these SOT/SP measurements were taken very close to the disk center, the returned magnetic parameters do not need to be converted to local solar coordinates. The synthetic profiles were convolved with the spectral PSF at the focal plane of the instrument. More details about the analysis of this SOT/SP data set can be found in \citet{Guglielmino:18}.

\textit{IRIS} acquired three data sets during the EFR evolution. Two of them consisted of single, very large, dense 400-step rasters (OBS3610108078), with simultaneous slit-jaw images (SJIs) composed of 1330, 1400, 2796, and 2832~\AA{} filtergrams. The first 400-step raster was made between 19:19 and 20:21~UT on April~13, the second between 02:09 and 03:11~UT on April~14.

\begin{figure*}[t]
	\centering
	\includegraphics[scale=0.65, clip, trim=25 30 45 240]{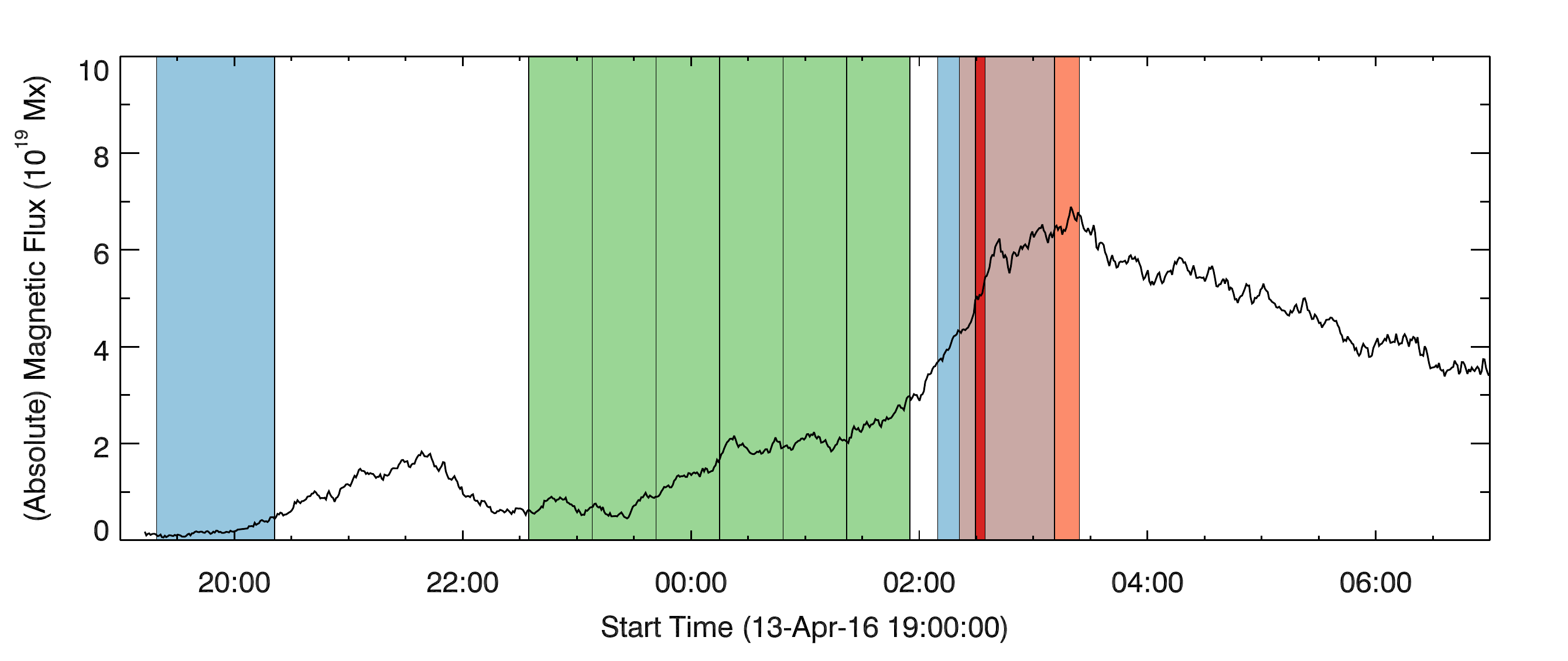}
	\caption{Evolution of the negative magnetic flux in the region within the solid box indicated in Figure~\ref{fig_context}, deduced from SDO/HMI measurements. The two \textit{IRIS} very large dense rasters are indicated with two blue strips. The \textit{IRIS} six large dense rasters relevant to this study are indicated with green strips. The \textit{Hinode} raster scan is represented by a red strip. A smaller, darker red strip indicates the time interval when the \textit{Hinode} SP slit passed over the EFR. The purple strips arise because of the superposition of blue and red strips, i.e., simultaneous observations by \textit{IRIS} and \textit{Hinode} satellites. \label{fig_flux}}
\end{figure*}

The third \textit{IRIS} data set, which is the most relevant to our study, is an observing sequence acquired between 22:34:43~UT on April~13 and 01:55:29~UT on April~14. It consisted of six large dense 64-step raster scans (OBS3610113456) for seven spectral windows. These included \ion{C}{2}~1334.5 and 1335.7~\AA{} lines, \ion{Si}{4} 1394 and 1402~\AA{} lines, 2814 and 2832~\AA{} bands, and \ion{Mg}{2} k~2796.4 and h~2803.5~\AA{} lines. Additionally, the spectral range of faint lines around \ion{O}{1} 1355.6~\AA{}, which comprises emission lines from hot ions such as \ion{Fe}{12} 1349~\AA{} and \ion{Fe}{21} 1354~\AA{}, was observed. The sequence had a 0\farcs33 step size and a 31.5~s step cadence, with a pixel size of 0\farcs35. The exposure time was initially 30~s, then on exposure 10 in raster 2 
it was reduced by the automatic exposure control to 9~s for the NUV channel and to 18~s for the FUV channels, respectively. The FoV of each scan was $22\farcs2 \times 128\farcs4$, as indicated in Figure~\ref{fig_context} with a dashed box. The raster cadence was about 33~min. Simultaneously, SJIs were acquired in the 1400 and 2796~\AA{} passbands, with a cadence of 63~s for consecutive frames in each passband, covering a FoV of $143\farcs7 \times 128\farcs4$. 

The \textit{IRIS} data were downloaded as level 2 products, already reduced by the instrument team. The version of the calibration processing IDL \textit{Solarsoft} routine (\textsc{iris\textunderscore{}prep}) applied to the data was 1.56. Intensities are given in normalized data-number units ($DN\,s^{-1}$), and were obtained by using the \textit{Solarsoft} routine \textsc{iris\textunderscore{}getwindata} \citep{Young:15}, which also provides the intensity uncertainties. The absolute velocity scale calibration for Doppler measurements was set by assuming zero velocity for the \ion{O}{1} 1355.60~\AA{}, \ion{S}{1} 1401.515~\AA{}, and \ion{Ni}{2} 2799.474~\AA{} cool lines, each present in one of the three \textit{IRIS} UV channels.

Full-disk data from the Atmospheric Imaging Assembly \citep[AIA;][]{Lemen:12} on board the \textit{SDO} satellite were also considered in the present work. In particular, we used images from the 1700, 1600, 304, 171, 335, and 131~\AA{} filters. The cadence of the SDO/AIA data is 12~s for the EUV channels and 24~s for the UV channels, respectively, with an image spatial scale of about 0\farcs6 per pixel.

In order to visualize the observing times of the high-resolution instruments used in this study, we plot in Figure~\ref{fig_flux} the evolution of the negative flux in the region of the EFR. We represent those observing time intervals as colored strips overplotted on the graph. 

\subsection{Co-alignment of the Observations}

The alignment between different instruments was obtained as follows. We used the SDO/HMI continuum filtergram taken at 22:35:25~UT, closest in time to the beginning of \textit{IRIS} observations, as a reference image. Consecutive SDO/HMI continuum filtergrams and simultaneous LOS magnetograms were aligned in sequence to the reference image, using the FoV shown in Figure~\ref{fig_context} as tracking for the cross-correlation algorithm. 

The SDO/AIA data were downloaded using the \textit{Solarsoft} cutout service, with the region of interest centered on the EFR. These SDO/AIA data are basically already aligned between them. Then, the 1700~\AA{} filtergrams closest in time to the SDO/HMI continuum filtergram at 22:35~UT were aligned to each other, by using the \textit{Solarsoft} mapping routines to take into account the different pixel sizes. Finally, the 1700~\AA{} channel was used as a reference for aligning the 1600~\AA{} channel and the remaining EUV channels.

To align the \textit{IRIS} observations to the \textit{SDO} data, we reconstructed continuum-like maps from the six scans, considering the integrated radiance in the 2832~\AA{} band, between 2831.9 and 2832.4~\AA{}. Each of these continuum-like maps was aligned through cross-correlation techniques with respect to the cospatial subFoV of the SDO/HMI continuum filtergram closest in time to the $32^{\mathrm{th}}$ \textit{IRIS} exposure, i.e., to the halfway time of each \textit{IRIS} scan, resampled to a pixel scale of 0\farcs33. The pores present in the subFoV were used as fiducial points.  

A similar procedure was used to align the \textit{Hinode} SOT/SP map to the \textit{SDO} data. We identified the acquisition time of the SOT/SP exposure relevant to the halfway point between the first (135) and the last (215) SP slit positions passing over the EFR. Then, the subFoV of the SOT/SP continuum map containing the EFR was aligned to the subFoV of the SDO/HMI continuum filtergram closest in time, resampled to a pixel size of 0\farcs32.

We estimate the accuracy of this procedure to be comparable to the pixel size of SDO/HMI data, i.e. $\pm 0\farcs5$.

\begin{figure}[t]
	\centering
	\includegraphics[scale=0.54, clip, trim=0 10 30 60]{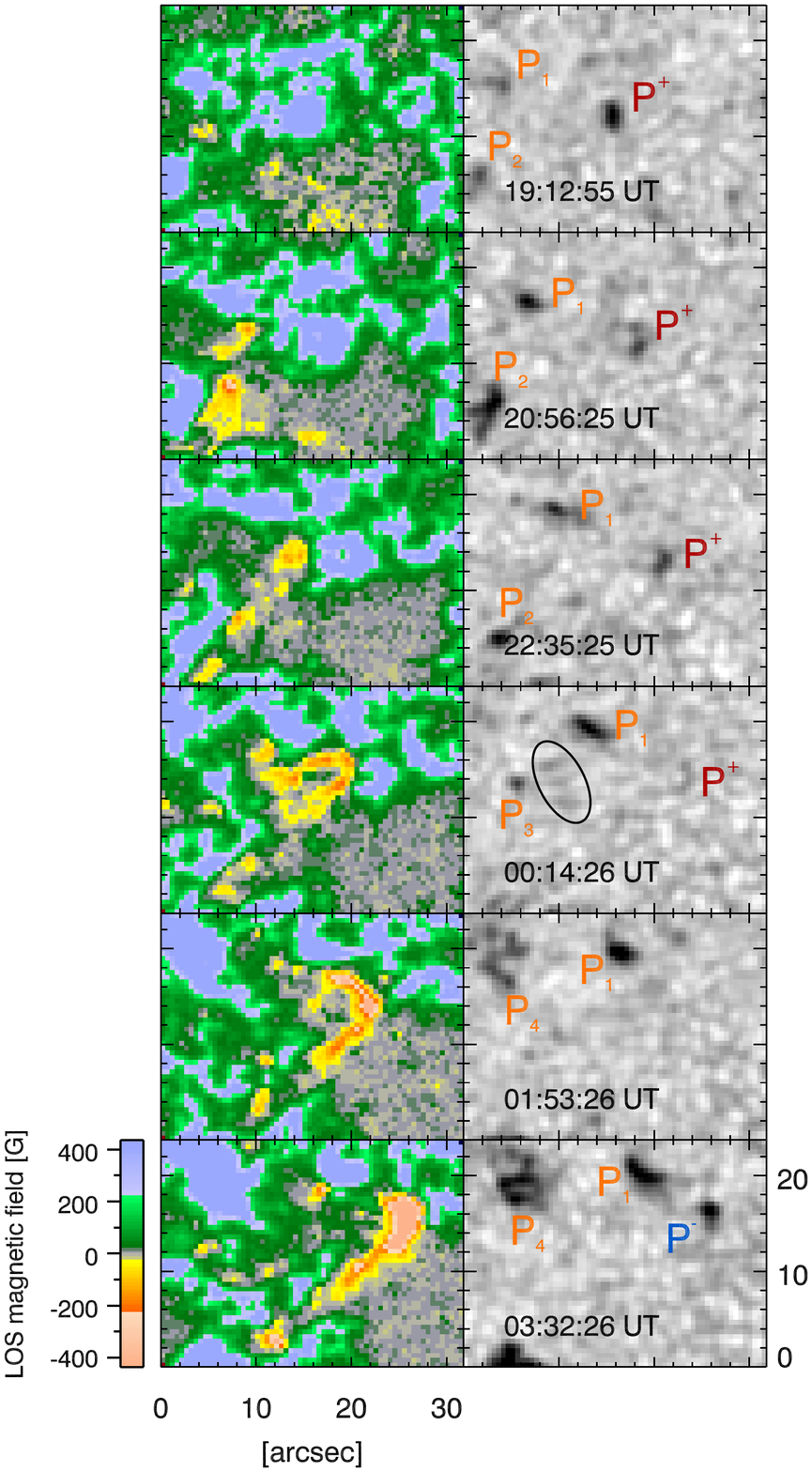}
	\caption{Evolution of the EFR as inferred from SDO/HMI observations. Left panels: LOS magnetograms. Right panels: Simultaneous normalized continuum intensity images. The pores are labelled according to their description in the main text. The oval indicates the location of the dark alignments. \\
	(An animation of this figure is available in the online Journal.) \label{fig_hmi}}
\end{figure}

\section{Results}

As already anticipated, the EFR that is analyzed in the present work appeared embedded within the following polarity of AR NOAA 12529. In the following subsections, we will report on the photospheric evolution of the EFR and we will describe the response of the upper atmospheric layers to flux emergence. Moreover, we will also analyze plasma properties in the EFR site deduced from \textit{IRIS} measurements.

\subsection{Photospheric evolution of the EFR}

Figure~\ref{fig_flux} presents the evolution of the amount of negative magnetic flux contained in the subFoV within the solid box indicated in Figure~\ref{fig_context}, deduced from SDO/HMI photospheric measurements. We have considered only those pixels with an (absolute) value larger than 25~G, about a factor of two of the noise level in SDO/HMI LOS measurements \citep[e.g.,][]{Hoeksema:14}. Moreover, we have taken into account the correction of the flux for the heliocentric angle. 

The graph in Figure~\ref{fig_flux} clearly shows that two flux emergence episodes occurred in the region of interest. The first occurred from 20:00 to 22:30~UT on April 13, with the flux peak at 21:45~UT, bringing into the photosphere a total flux content of about $2 \times 10^{19} \,\mathrm{Mx}$. The rise phase of the second flux emergence episode took place from 23:30 on April 13 to 03:30~UT on April 14. At this time, the total amount of emerged flux peaked at $\approx 7 \times 10^{19} \,\mathrm{Mx}$. The decay phase lasted until the end of the observing time considered for SDO observations.

In Figure~\ref{fig_hmi} we report the history of the EFR as inferred from SDO/HMI observations. LOS magnetograms (\textit{left panels}) reveal that, at the beginning of the observations, in the subFoV there was essentially a unipolar plage (19:12~UT). The first flux emergence episode occurred in the region at X=$[4\arcsec, 8\arcsec]$, Y=$[2\arcsec, 14\arcsec]$ (see, e.g., the magnetogram at 20:56~UT). Later, the second flux emergence event began in the region comprised between X=$[8\arcsec, 16\arcsec]$, Y=$[6\arcsec, 16\arcsec]$ at 22:35~UT. The negative, emerging polarity was clearly seen at X=$14\arcsec$, Y=$14\arcsec$, close to a pre-existing flux concentration near the center of the subFoV, which was conversely characterized by positive polarity. At 00:14~UT one can observe the growing negative polarity of the EFR, with a semi-circular footpoint. The positive polarity was located in the region at X=$[6\arcsec, 10\arcsec]$,  Y=$[8\arcsec, 14\arcsec]$. At 01:53~UT, the positive polarity of the EFR already merged into the diffuse ambient field with the same polarity. Interestingly in the meantime, the positive flux concentration near the center of the subFoV, which was in contact to the growing, negative polarity of the EFR, became smaller until it totally disappeared, as clearly seen in the online movie. Finally, at 03:32~UT, one can note a flux concentration newly formed by the accumulation of the negative polarity flux carried by the EFR at X=$[22\arcsec, 28\arcsec]$, Y=$[12\arcsec, 18\arcsec]$.

The simultaneous continuum intensity filtergrams (Figure~\ref{fig_hmi}, \textit{right panels}) show the evolution of the corresponding phenomena in the photosphere. At the beginning of the observations (19:12~UT), there was a pore (P$^{+}$) at the center of the subFoV, cospatial to a flux concentration with positive polarity. The two pores at the left edge of the subFoV, P$_{1}$ and P$_{2}$, corresponding to strong flux concentrations with positive polarity as well, were growing while moving to the west, as clearly visible at 20:56~UT. Meanwhile, the pore P$^{+}$ near the center was shrinking (22:35~UT) as long as the positive flux patch cospatial to it came in contact with the emerging, negative polarity of the EFR. The pore P$^{+}$ was only faintly distinguishable at 00:14~UT, eventually disappearing at 01:53~UT. P$_{1}$ continued growing, whereas P$_{2}$ disappeared at 00:14 UT. At that time, small-scale dark elongated features, i.e., alignments in the dark intergranular lanes, were observed in the area at X=$[8\arcsec, 14\arcsec]$, Y=$[10\arcsec, 16\arcsec]$, as indicated by an oval enclosing them in the image. These features seem to converge on the small pore P$_{3}$ appeared at X=$6\arcsec$, Y=$14\arcsec$ that was no longer visible in the following frames. The pore P$_{4}$ entered from the left into the subFoV at 01:53~UT. P$_{4}$ became darker and slightly larger by 03:32~UT. At the same time, a new pore (P$^{-}$) appeared at the right side of the subFoV at X=$26\arcsec$, Y=$16\arcsec$, cospatial with the negative flux patch formed by the accumulation of the negative polarity flux carried by the EFR.

\begin{figure}[t]
	\centering
	\includegraphics[scale=0.75, clip, trim=30 60 250 55]{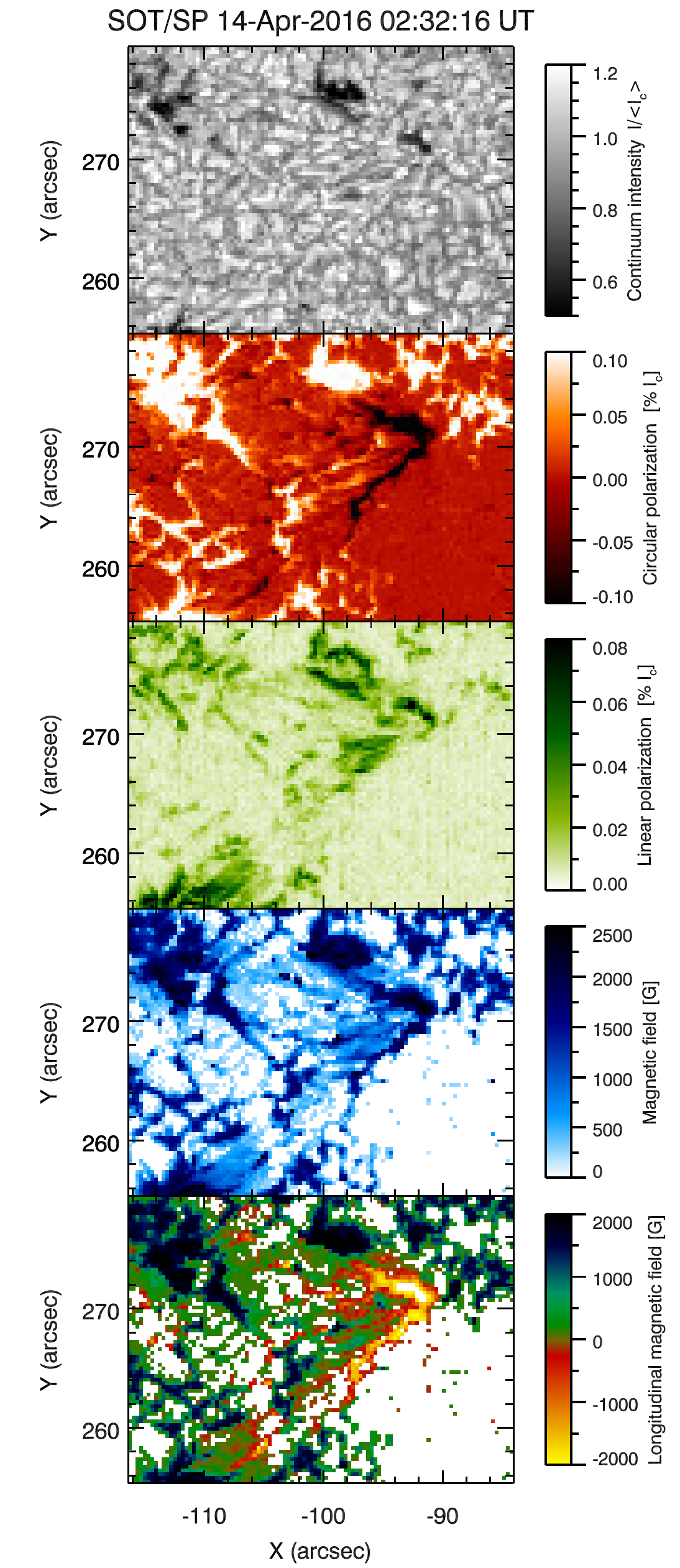}
	\caption{{\tiny Maps of normalized continuum intensity (\textit{top panel}), circular and linear polarization (\textit{second and third panels}), magnetic field strength (\textit{fourth panel}), and longitudinal field component (\textit{bottom panel}), deduced from \textit{Hinode} SOT/SP observations acquired between 02:29 and 02:35~UT. The white background in the maps of the magnetic field strength and of the longitudinal field component represents pixels with total polarization $<1.5\%$, which are not considered in the inversion.} \label{fig_sot}}
\end{figure}

In summary, SDO/HMI observed that
	\begin{itemize}
		\setlength\itemsep{0em}
		\item an EFR emerged in a unipolar (positive) plage;
		\item a pre-existing flux concentration, corresponding to a pore (P$^{+}$), became 
		smaller and finally disappeared while being in contact with the growing polarity of the EFR with opposite (negative) polarity;
		\item the accumulation of the negative polarity flux brought into the photosphere by the EFR led to the formation of a new pore (P$^{-}$).
\end{itemize}

Later, at a time close to the peak of flux, the slit of SOT/SP passed over the EFR, from 02:29 to 02:35~UT. 
This data set is not directly relevant to the \textit{IRIS} observations during the emerging phase of the EFR. Though, even if only in a cursory way, it is worth giving here a hint about the photospheric configuration retrieved by these measurements, as fully representative of the late stages of the EFR, almost fully developed.

\begin{figure*}[t]
	\centering
	\includegraphics[scale=0.75, clip, trim=40 40 190 20]{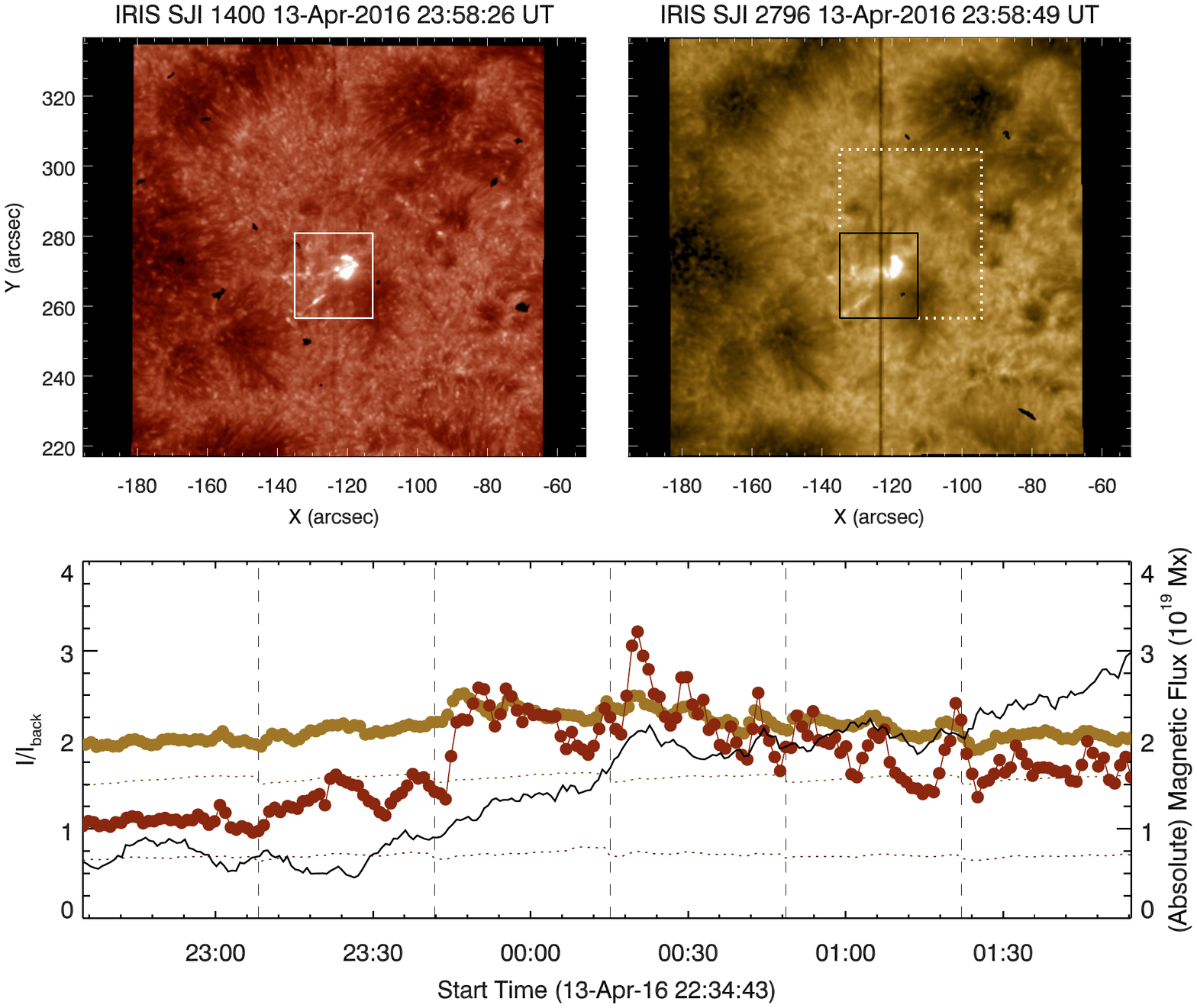}
	\caption{\textit{Top panels:} \textit{IRIS} SJI 1400~\AA{} and 2796~\AA{} co-temporal images that outline chromospheric and TR features, relevant to the third raster scan of the analyzed sequence. The solid white (black) box marks the common subFoV between the area scanned by the \textit{IRIS} slit and the region where the EFR is observed in the SDO/HMI filtergrams (solid box in Figure~\ref{fig_context}). This subFoV is used for the more detailed investigation. The dashed line box indicates the subFoV used to study the jet analyzed in Figure~\ref{fig_jets}. \textit{Bottom panel}: intensity variation with time as obtained from the \textit{IRIS} 1400~\AA{} (reddish colour) and 2796~\AA{} (yellowish colour) passbands in the above mentioned area, indicated with a solid box. The beginning and end times of the \textit{IRIS} rasters are indicated with dashed vertical lines. The solid black line indicates the flux trend of the negative polarity of the EFR during the \textit{IRIS} scans. \\
	(An animation of this figure is available in the online Journal.) \label{fig_irisfov}}
\end{figure*}

Figure~\ref{fig_sot} illustrates the maps of continuum intensity (\textit{top panel}), circular and linear polarization (\textit{second and third panels}). The maps of the magnetic field strength and of the longitudinal field component, deduced from the inversion of the SOT/SP spectra, are also displayed (\textit{fourth and bottom panels}). The higher spatial resolution of SOT/SP data makes visible the fine structure of the EFR. The serpentine field, associated with a mixed polarity pattern, was clearly distinguishable in the emergence zone, i.e., the region between the emerging polarities of the EFR, (\textit{Solar} X=$[-98\arcsec, -92\arcsec]$, \textit{Solar} Y=$[266\arcsec, 270\arcsec]$), both in the circular polarization map and in the longitudinal field map. In this area, we have also found elongated granulation in the continuum map and enhanced linear polarization signals. The average value of $B$ in the emergence zone is about 1000~G. In the polarities of the EFR, $B$ is $\approx 2000\,\mathrm{G}$ in the compact negative polarity, which was forming a pore, and 1500~G in the more diffuse, positive polarity. 

Note that the total amount of negative flux contained within the subFoV deduced from the SOT/SP measurements is $\sim 2.4 \times 10^{20} \,\mathrm{Mx}$, which is a factor of three larger than the value estimated from SDO/HMI observations at the same time. This is due to the differences in spatial resolution and wavelength sampling between SDO/HMI and \textit{Hinode} SOT/SP, so that the SDO/HMI flux density generally is lower than that reported by SOT/SP, which is more accurate \citep{Hoeksema:14}. For instance, while studying the increment of the magnetic field strength during the solar flare trigger process, \citet{Bamba:14} found a difference of about a factor of two in the magnetic field strength inferred by the two instruments, using a Milne-Eddington inversion for the SOT/SP data. 

\subsection{Response of the upper atmospheric layers}

A visual inspection of the \textit{IRIS} SJIs, acquired between 23:34~UT on April~13 and 01:55~UT on April~14, discloses intense, repeated brightenings occurring near the center of the FoV of these images (Figure~\ref{fig_irisfov}, \textit{top panels}). In particular, we see a compact UV burst, which had a size of about $ 7\arcsec \times 7\arcsec$ and lasted throughout the observing sequence, i.e., more than three hours, slightly changing in size and intensity with time (see also the online movie). 

The spatial configuration of the UV burst deduced from the 1400 SJIs appears to consist of many sub-structures that are transitory and very dynamic. However, it is difficult to detail the fine structure owing to the fact that images show saturation.

\begin{figure*}[t]
	\centering
	\includegraphics[scale=0.625, clip, trim=0 5 25 0]{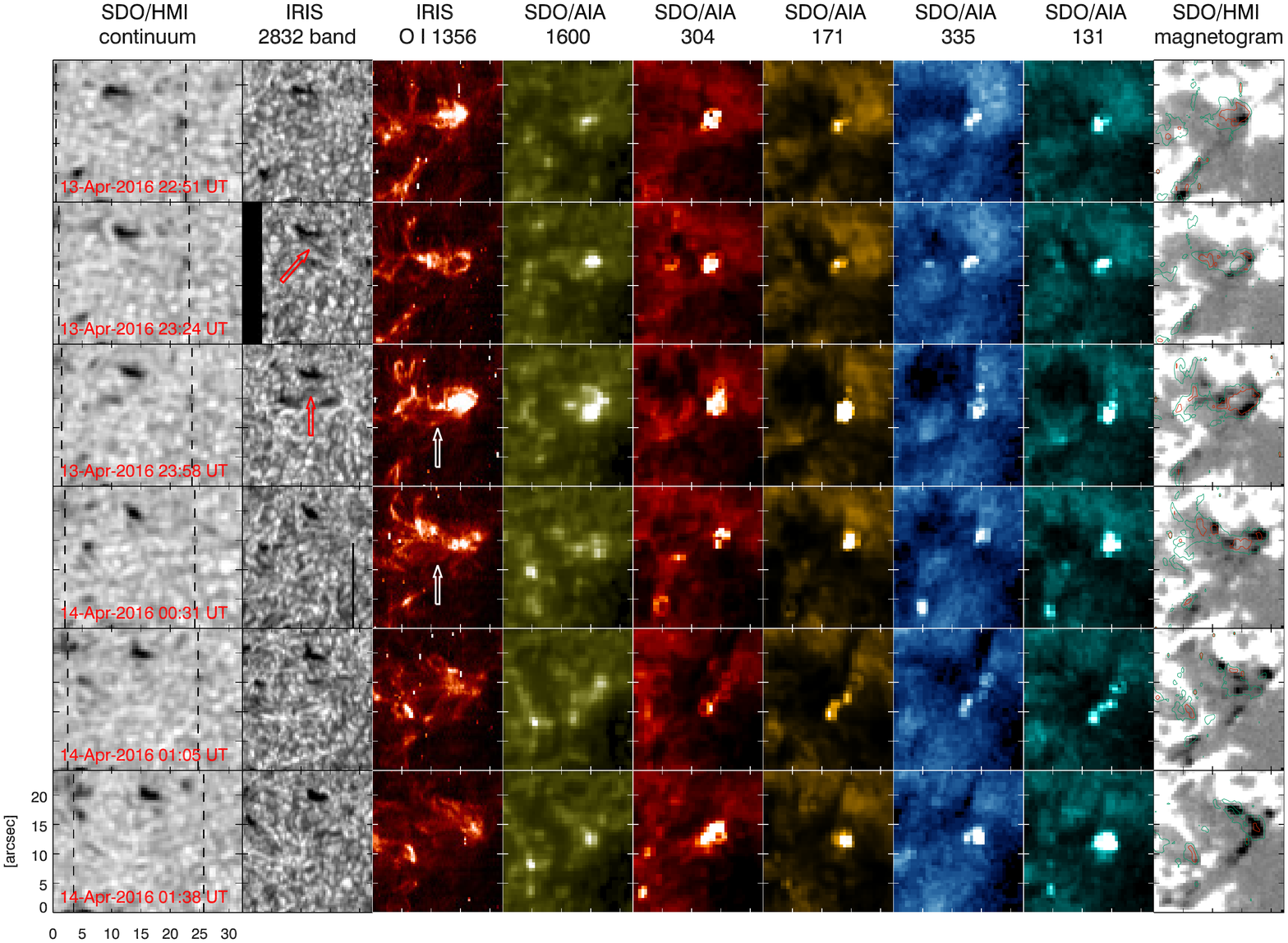}
	\caption{Synoptic view of the evolution of the EFR at different atmospheric layers, during the \textit{IRIS} observing sequence. From left to right, each row shows the SDO/HMI continuum map, the \textit{IRIS} reconstructed maps of the radiance in the 2832~\AA{} band and in the \ion{O}{1} 1355.6~\AA{} line, and the SDO/AIA cospatial maps for the selected channels. Finally, the last column refers to the simultaneous SDO/HMI LOS magnetogram, with overplotted contours of the \ion{O}{1} 1355.6~\AA{} radiance. The time of both SDO/HMI and SDO/AIA filtergrams is the closest to the halfway time of each \textit{IRIS} scan. \label{fig_synoptic}}
\end{figure*}

The UV lightcurves derived for the common FoV between the area scanned by the \textit{IRIS} slit and the region where the EFR appeared in the SDO/HMI filtergrams (solid boxes) have been plotted in Figure~\ref{fig_irisfov} (\textit{bottom panel}). They illustrate that the larger variation in brightness occurred in the 1400~\AA{} passband (reddish colour), with a strong rise during the third scan and a peak during the fourth one, followed by several peaks that decrease in amplitude with a bursty behaviour. Conversely, the intensity in the 2796~\AA{} passband (yellowish colour) appears to be more steady, albeit an increase was detected in third scan, followed again by a number of peaks decreasing in amplitude. However, the trend is smoother than in the 1400~\AA{} passband. 

For comparison, we have also plotted the trend of the negative magnetic flux emerging in the EFR relevant to the time interval of these \textit{IRIS} observations (solid black line). This graph indicates that the brightness enhancements occurred during the rise phase of the negative (emerging) flux. Noticeably, many intensity peaks took place a few minutes after steeper flux increments.

Figure~\ref{fig_synoptic} pictures the evolution of the EFR at several atmospheric layers during the \textit{IRIS} observing sequence, with a $\sim 33$~minutes cadence. 

The SDO/HMI continuum maps (Figure~\ref{fig_synoptic}, first column) report the photospheric evolution of the EFR for a time comprised in the interval between the third and the fifth rows of Figure~\ref{fig_hmi}. These maps can be easily compared with the \textit{IRIS} reconstructed maps of the radiance in the 2832~\AA{} band (second column), which refers to the photosphere as well. Note that the dashed vertical lines in the SDO/HMI continuum maps indicate the edge of the area scanned by the \textit{IRIS} slit, used as subFoV in the other maps. In the \textit{IRIS} 2832~\AA{} maps, thanks to the higher spatial resolution compared to the SDO/HMI data, the dark alignments between the pores, cospatial to the horizontal emerging field of the EFR, are remarkably more visible. Specifically, they appeared in the region at X=$[10\arcsec, 15\arcsec]$, Y=$[15\arcsec, 20\arcsec]$ at 23:24~UT and at X=$[7\arcsec, 18\arcsec]$, Y=$[15\arcsec, 20\arcsec]$ at 23:58~UT, as indicated with red arrows in the figure. As for the rest of the time, the evolution is in agreement with the description already given for Figure~\ref{fig_hmi}.

\begin{figure}[t]
	\centering
	\includegraphics[scale=0.575, clip, trim=15 25 115 135]{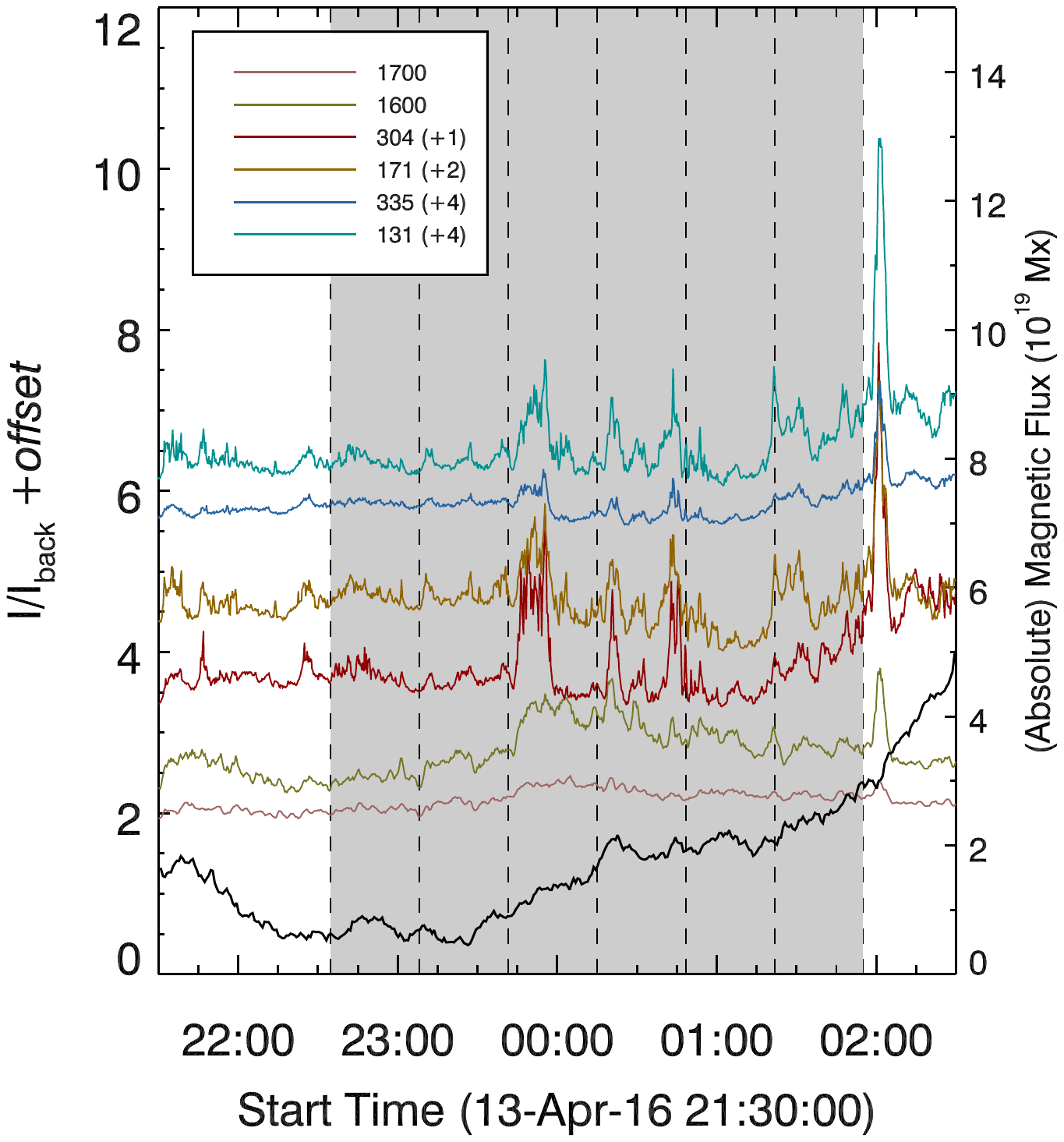}
	\caption{Lightcurves for the UV and EUV channels SDO/AIA in the subFoV shown in Figure~\ref{fig_synoptic}, deduced during the evolution of the EFR. Note the offsets of the lightcurves along the \textit{y} direction, to enhance their visibility. The grey-shaded area indicates the time interval of the \textit{IRIS} rasters. The beginning and end times of each \textit{IRIS} raster are indicated with dashed vertical lines. For comparison, we also plot the trend of the negative magnetic flux in the EFR (black line), already presented in Figure~\ref{fig_flux}.  \label{fig_aiacurves}}
\end{figure}

The response at the chromospheric level to flux emergence has been investigated by analyzing the \textit{IRIS} reconstructed maps of the radiance in the \ion{O}{1} 1355.6~\AA{} line (Figure~\ref{fig_synoptic}, third column) and the cospatial SDO/AIA filtergrams in the 1600~\AA{} and 304~\AA{} passbands (Figure~\ref{fig_synoptic}, fourth and fifth columns). The \ion{O}{1} 1355.6~\AA{} radiance has been deduced from a Gaussian fit to the line, using the \textit{Solarsoft} routine \textsc{eis\textunderscore{}auto\textunderscore{}fit}. This procedure allows the users to fit lines simultaneously with multiple independent Gaussians, in order to remove possible blending with nearby lines \citep{Young:09}. The \ion{O}{1} 1355.6~\AA{} line is an excellent probe of the middle chromosphere \citep{Lin:15}. In this layer, we observe the presence of a bundle of arches above the EFR, which constitute an arch filament system \citep[AFS; e.g.,][]{Bruzek:80,Spadaro:04,Zuccarello:05}. Intensity enhancements took place to the West edge of this structure, the strongest brightenings being observed during the first and third \textit{IRIS} scans. They appeared as a perfect counterpart of the UV burst seen in the 1400~\AA{} and 2796~\AA{} SJIs (see Figure~\ref{fig_irisfov}). However, the radiance in the \ion{O}{1} 1355.6~\AA{} line presented a decline in the following scans. 

Similarly, the 1600~\AA{} channel exhibited an intensity increase that reached its maximum in the region cospatial to the UV burst at 23:58~UT, during the third \textit{IRIS} scan, then faded away. In comparison, the 304~\AA{} channel also shows a brightness enhancement from 22:51~UT to 23:58~UT, albeit it is significantly smoother than in the 1600~\AA{} channel. Later, we see a decrease until 01:05~UT, and a new increment of emission at 01:38~UT, that was not seen at the lower chromospheric levels. 

The contrasting trend of the 1600~\AA{} and 304~\AA{} channels is also reflected in the lightcurves plotted in Figure~\ref{fig_aiacurves}. This graph displays the lightcurves for the SDO/AIA channels analyzed in the present work, including the 1700~\AA{} channel as well. Images through this filter have not been shown in Figure~\ref{fig_synoptic}, as they refer to the upper photosphere and are rather similar to those acquired in the SDO/AIA 1600~\AA{} passband. The lightcurves have been computed for the subFoV of SDO/AIA filtergrams shown in Figure~\ref{fig_synoptic}, using a nearby quiet-Sun area as a reference for the background intensity. The 1700~\AA{} channel exhibits a slight intensity increase during the \textit{IRIS} observing time, with a peak between the third and fourth \textit{IRIS} scans. The 1600~\AA{} and 304~\AA{} passbands, instead, show a conspicuous brightness enhancement at the beginning of the third \textit{IRIS} scan. Then, the 1600~\AA{} intensity had a slow decrease until the end of \textit{IRIS} observations, whereas the 304~\AA{} passband showed a bursty trend with a final rise during the sixth \textit{IRIS} scan. It is worth mentioning that the strongest intensity peak found in all the SDO/AIA channels was unfortunately observed just a few minutes after the end of the \textit{IRIS} observations (see Figure~\ref{fig_aiacurves} slightly after 02:00~UT).

At coronal level, brightenings were observed in the three SDO/AIA channels at 171~\AA{}, 335~\AA{}, and 131~\AA{} (Figure~\ref{fig_synoptic}, sixth, seventh, and eighth columns). These passbands refer to the low ($T=0.7$~MK), middle ($T \leq 2.5$~MK), and high ($T=10$~MK) corona, respectively. However, the SDO/AIA 131~\AA{} has emission contributions from two dominant ions, which are formed at completely different temperatures: \ion{Fe}{8} at $0.6$~MK and \ion{Fe}{20} at $10$~MK. In non-flaring conditions, emission from the very hot \ion{Fe}{20} ion is usually negligible \citep{Sykora:11}. A contribution of the channels emission due to the continuum enhancement is expected to be negligible, particularly through the 171~\AA{} filter, which is relatively uncontaminated. Actually, \citet{ODwyer:10} found that TR and continuum emission in the 171~\AA{} channel is at least a factor 100 weaker than the coronal emission for AR conditions. The brightness enhancements found at the location of the UV burst in the AIA passband are a factor 10 brighter than a nearby background region, whereas in the TR we observed about a factor 100 emission increase in the \ion{Si}{4} 1402~\AA{} line with respect to the background (see Sect.~3.3). Therefore, if the TR and continuum were enhanced by the same factor as \ion{Si}{4} 1402~\AA{} line, then it could not explain the factor 10 brightening in the 171~\AA{} channel. Thus, the contamination of the SDO/AIA passbands from the continuum emission appears to be not significant for these bursts. Note also that, albeit not shown here, similar brightness enhancements were also observed at the same locations in the SDO/AIA 193~\AA{} channel, which is dominated by \ion{Fe}{12} lines. This provides a further evidence of genuine coronal emission.

\begin{figure*}[t]
	\centering
	\includegraphics[scale=0.625, clip, trim=0 0 25 0]{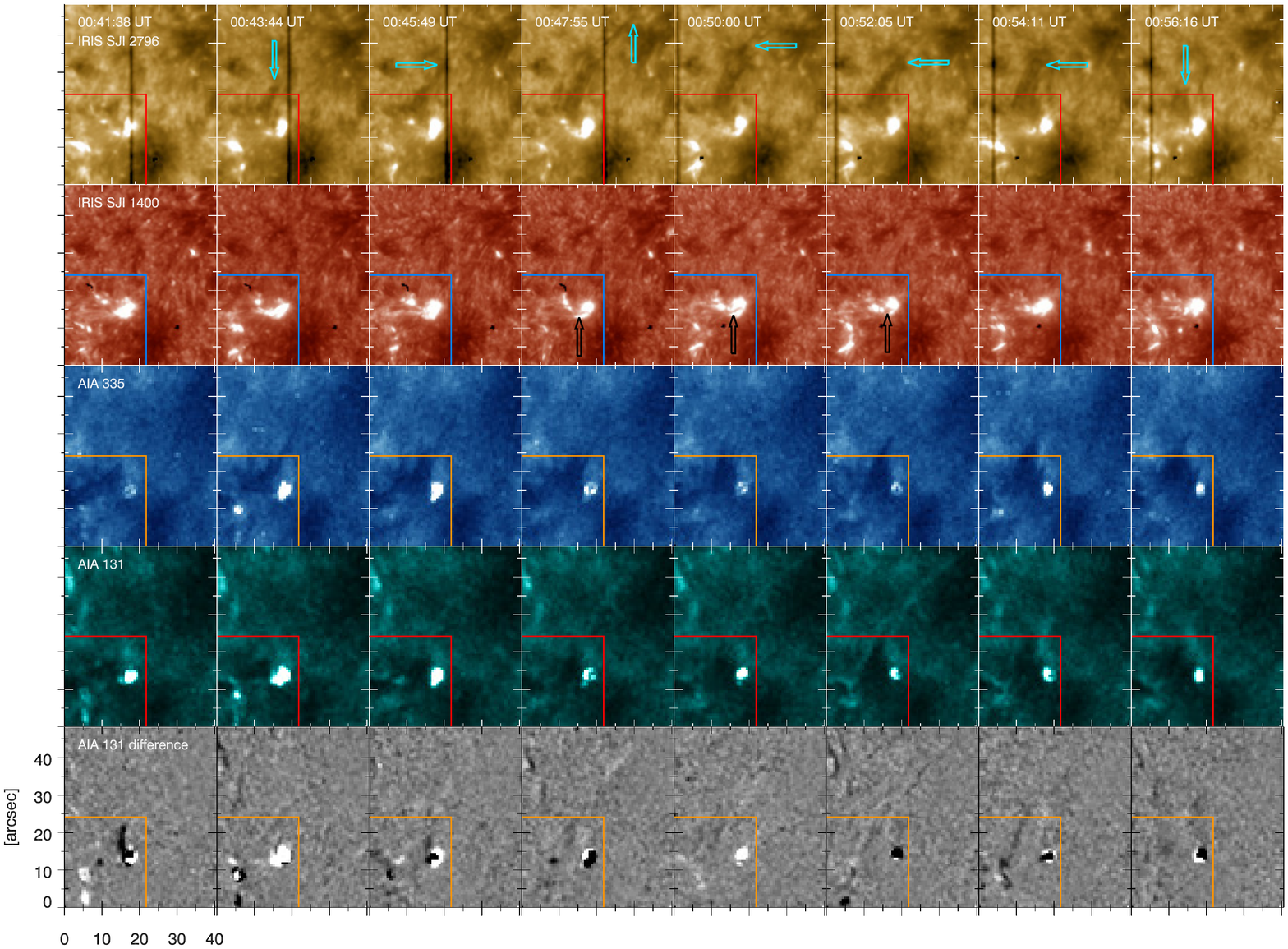}
	\caption{Sequence of \textit{IRIS} SJIs and SDO/AIA filtergrams showing the evolution of a plasma ejection at different atmospheric heights, with a $\sim 2$ minutes cadence. For comparison, the colored box in all the frames marks the subFoV shown in Figure~\ref{fig_synoptic}. The blue arrows in the first row indicate the end-point of the surge-like ejection. \\
	(An animation of this figure is available in the online Journal.) \label{fig_jets}}
\end{figure*}

As one can see in Figure~\ref{fig_synoptic}, the evolution of the brightening site in these channels resembles that found in the 304~\AA{} channel. The brightness enhancement was observed throughout the \textit{IRIS} observing time. A first, abrupt intensity peak occurred during the third \textit{IRIS} scan, visible in the corresponding EUV lightcurves displayed in Figure~\ref{fig_aiacurves} as well. Then, a phase with a bursty trend followed, until a new intensity increase took place during the sixth \textit{IRIS} scan, with the strongest peak seen soon after the end of the \textit{IRIS} observations, as already noted.

The last column of Figure~\ref{fig_synoptic} shows the SDO/HMI LOS magnetogram simultaneous to the continuum filtergram displayed in the first column of the same Figure, with the subFoV of \textit{IRIS} rasters. Moreover, contours at 85\% (blue) and 95\% (red) of the maximum radiance in the \ion{O}{1} 1355.6~\AA{} line have been overplotted on the maps. It is easily seen that the brightening site was essentially localized in the interface region between the emerging, negative polarity of the EFR (black patches) and the pre-existing, positive polarity of the ambient field (white patches) where the EFR was embedded. However, one can notice a slight displacement of the burst toward the East direction with respect to the polarity inversion line.

\begin{figure*}[t]
	\centering
	\includegraphics[scale=0.325, clip, trim=100 112 200 120]{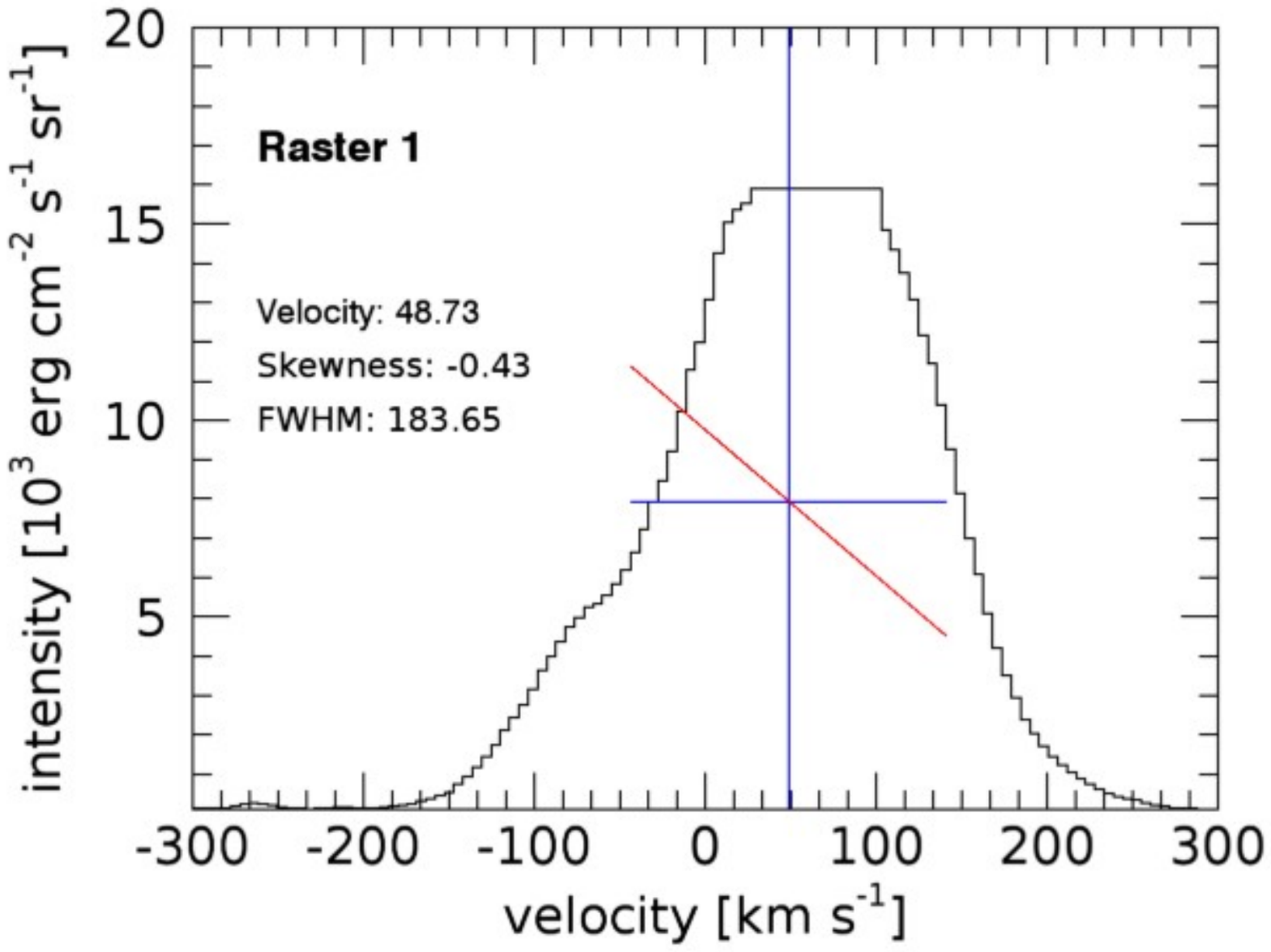}%
	\includegraphics[scale=0.325, clip, trim=145 112 200 120]{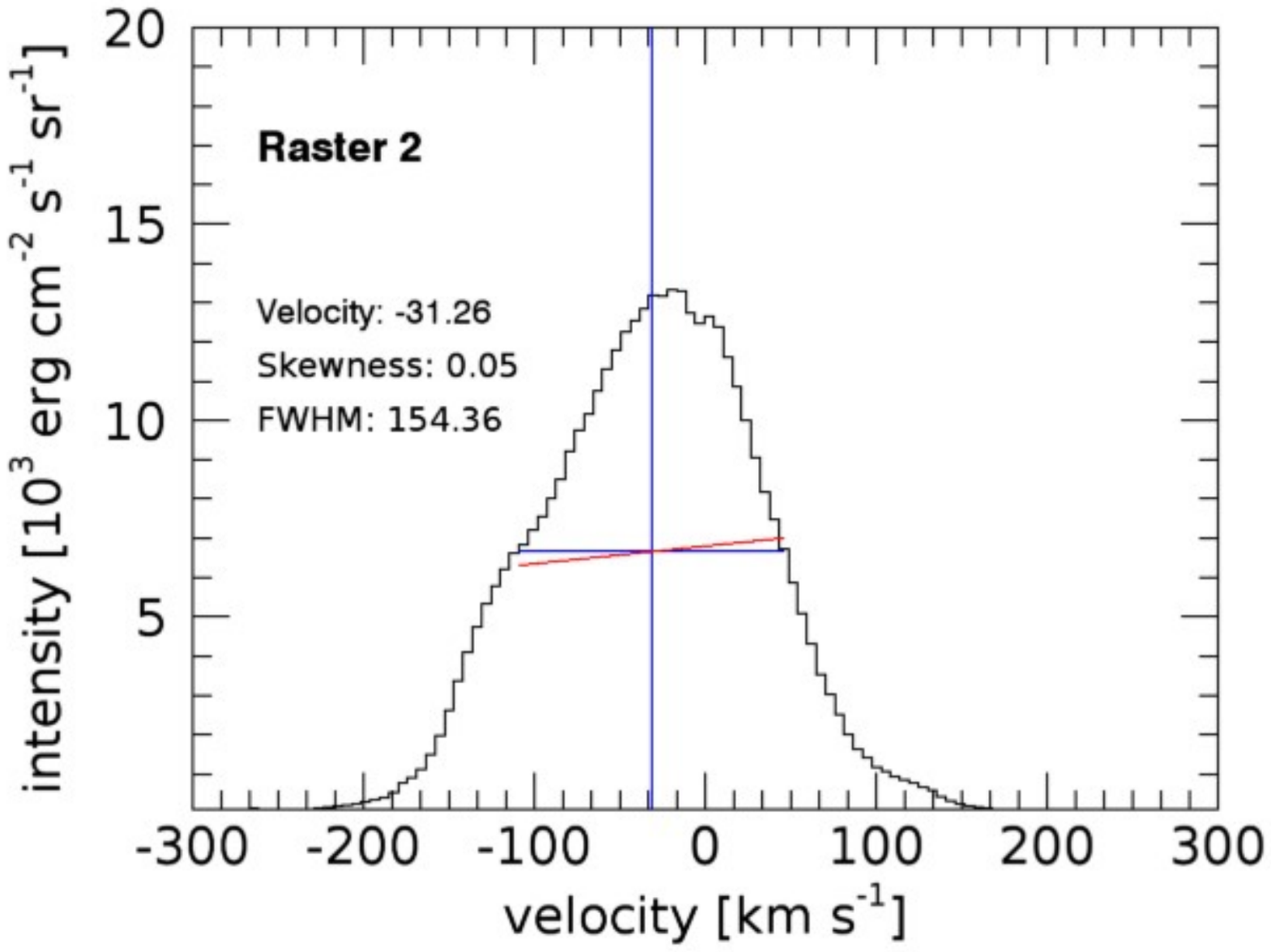}%
	\includegraphics[scale=0.325, clip, trim=145 112 200 120]{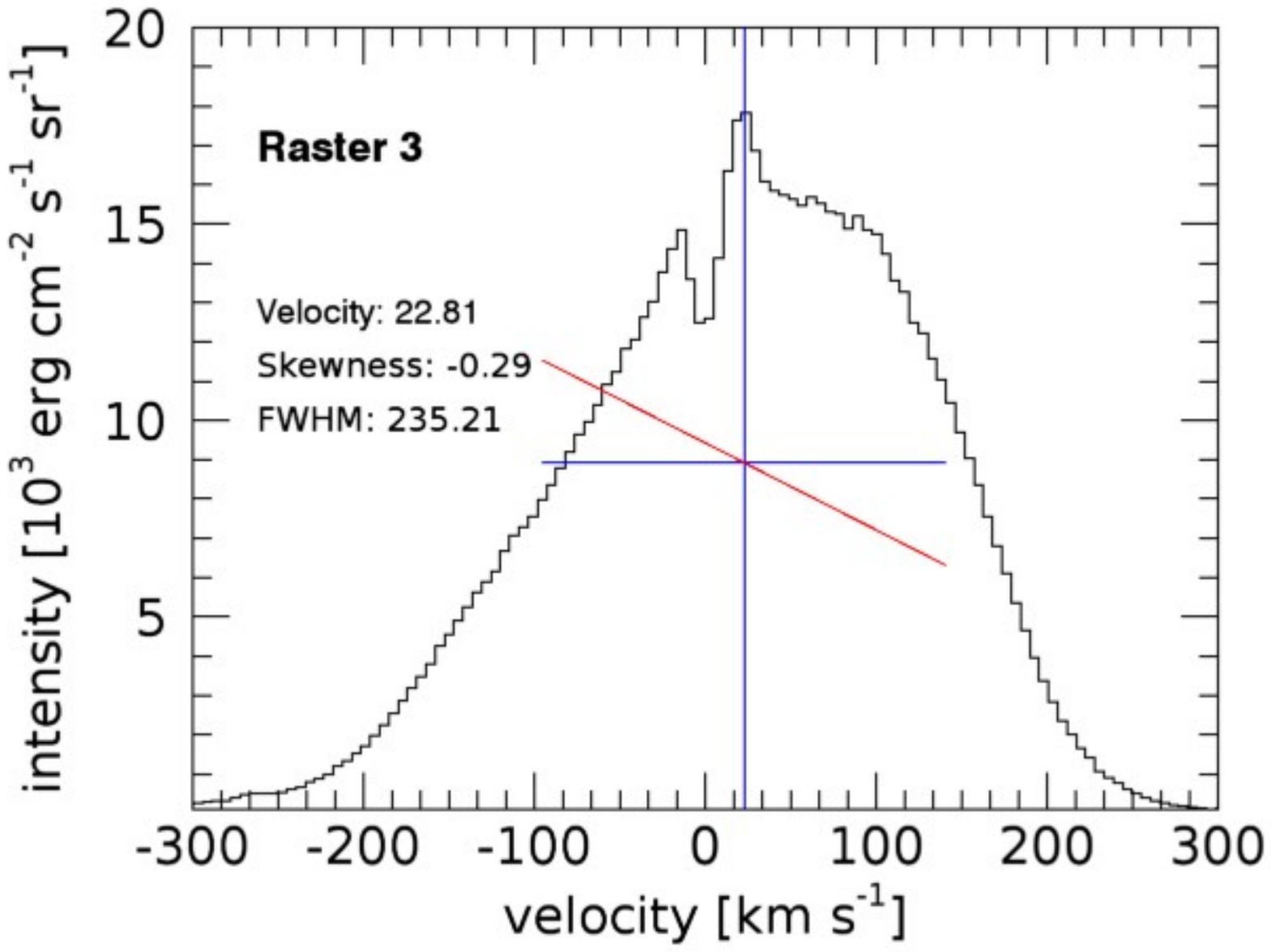}
	\includegraphics[scale=0.325, clip, trim=100 107 200 120]{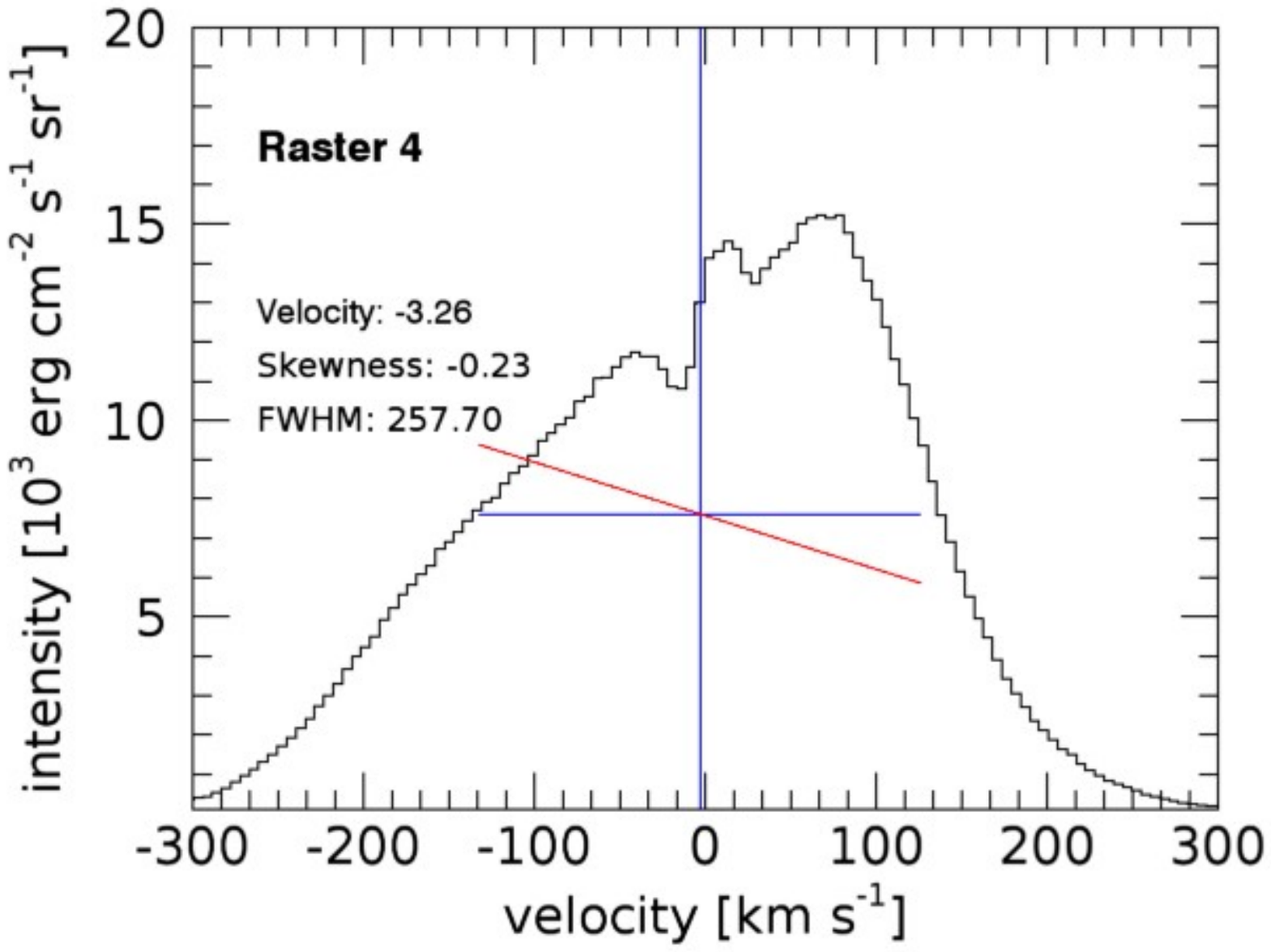}%
	\includegraphics[scale=0.325, clip, trim=145 107 200 120]{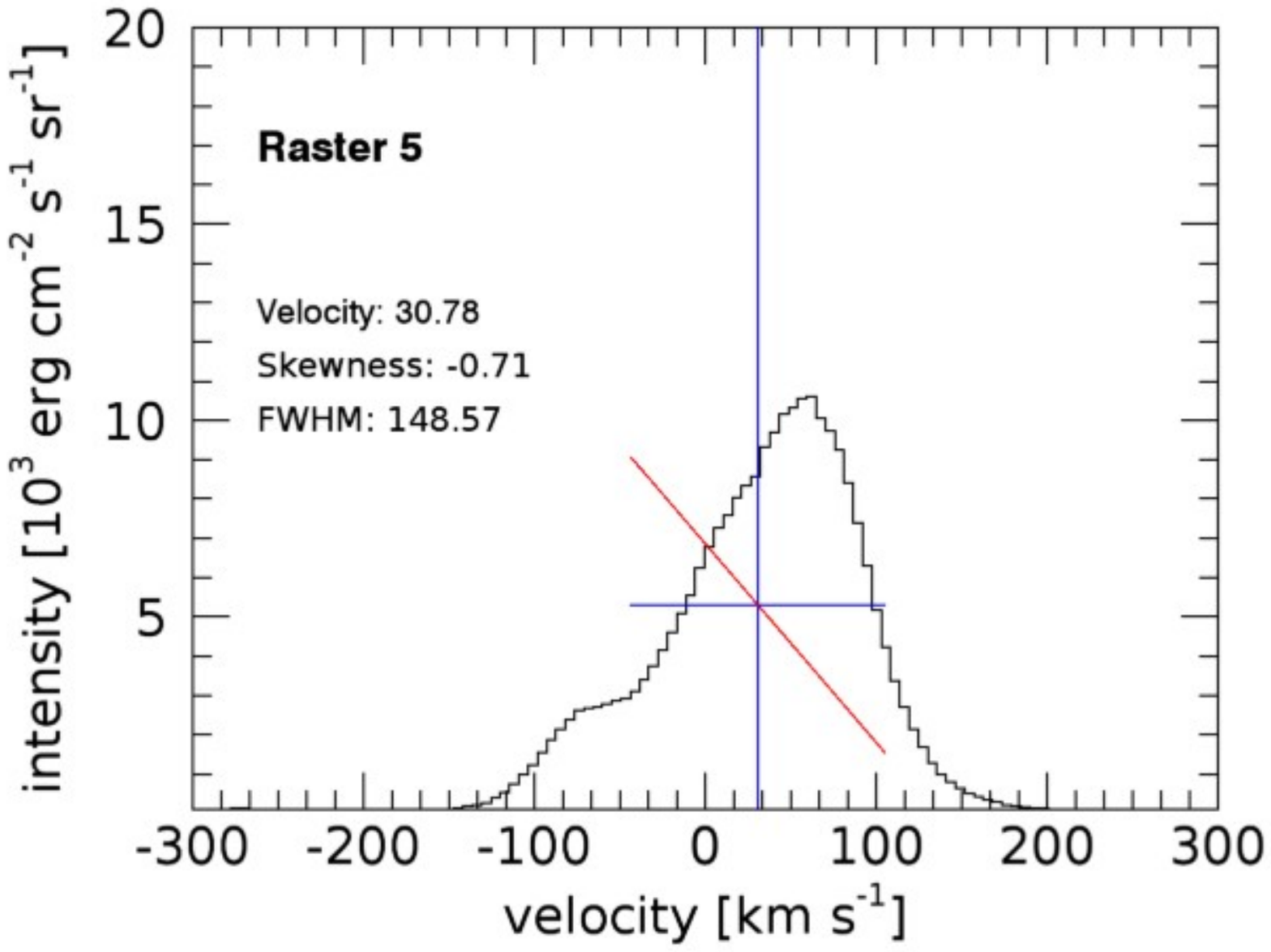}%
	\includegraphics[scale=0.325, clip, trim=145 107 200 120]{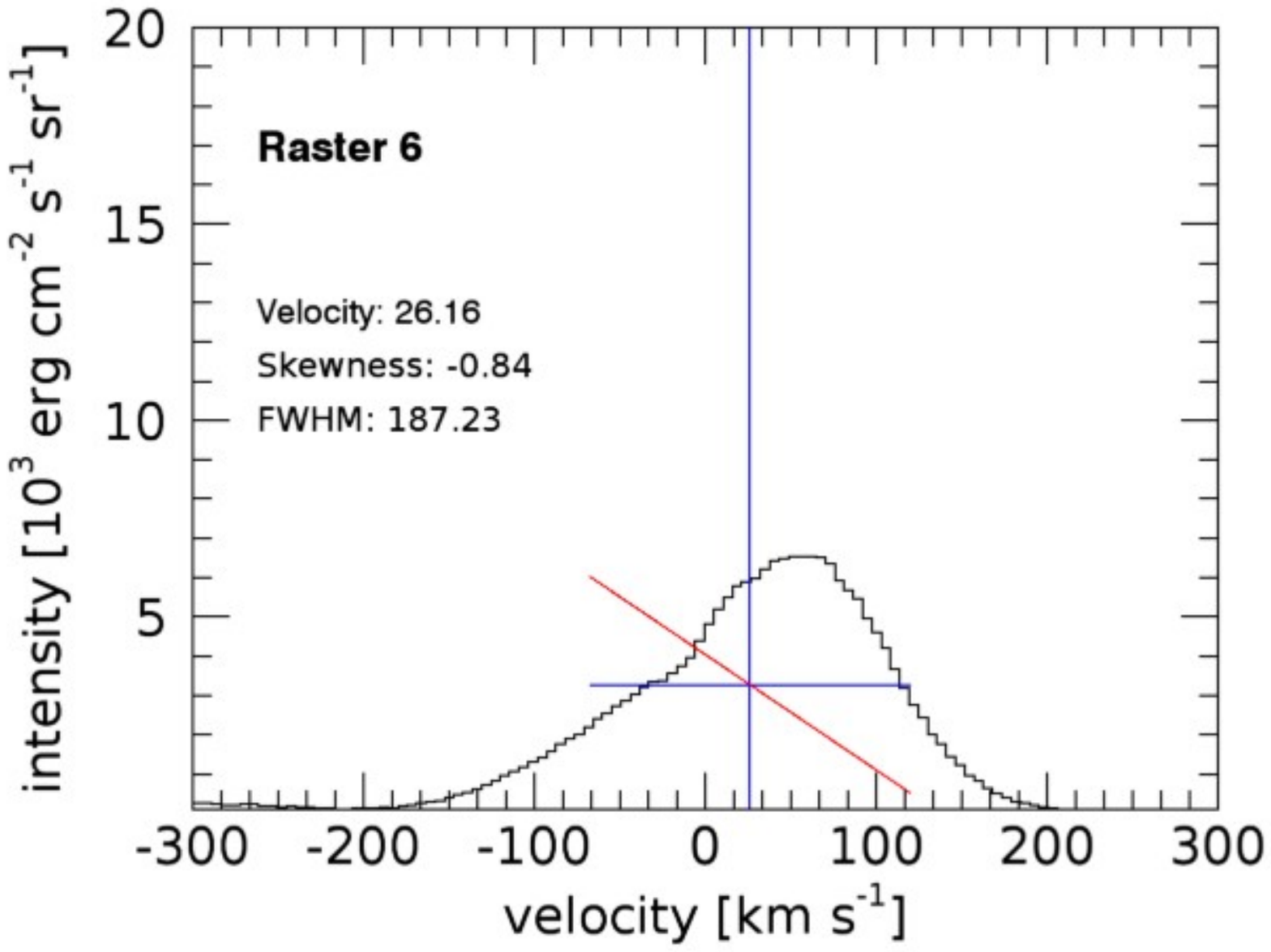}
	\caption{Spectral line profiles of the \ion{Si}{4} 1402~\AA{} line in the pixels with maximum integrated intensity, in each of the six \textit{IRIS} raster scans. The blue and red segments visually represent the width and the skewness that come from the moment analysis, respectively. The blue segment is proportional to the FWHM. A negative (positive) slope of the red segment indicates the profile is skewed to the blue (red). A steeper slope indicates a larger amount of skewness.  \label{fig_irisbrightest}}
\end{figure*}

Another phenomenon, i.e., plasma ejections, occurred during the evolution of the EFR. In the online movie relative to the 2796~\AA{} passband in Figure~\ref{fig_irisfov}, one can recognize a number of surge-like events departing in the region toward the East with respect to the UV burst, cospatial with the AFS. These ejections appeared as filamentary elongating structures, darker than their surroundings.

Figure~\ref{fig_jets} illustrates the development of the most conspicuous among these surge-like ejections observed by \textit{IRIS}. The panels of Figure~\ref{fig_jets} image the area of SJIs framed with a dashed box in Figure~\ref{fig_irisfov} (\textit{top-right panel}). In particular, we see the ejection in the first row in the 2796~\AA{} SJIs, as indicated by blue arrows. This surge-like event had a lifetime of about 15~minutes, reaching its maximum extension after about 6~minutes since its beginning. The apparent maximum length is $\sim 27\arcsec$ ($19\,\mathrm{Mm}$), which corresponds to an average projected horizontal velocity of about $50 \,\mathrm{km\,s}^{-1}$. There is no counterpart of the ejection in the 1400~\AA{} SJIs (Figure~\ref{fig_jets}, second row), except for a darker, fluffy-like region that slightly clouds the UV burst at its left edge, in correspondence with the base of the surge-like structure (see black arrows).

\begin{figure*}[t]
	\centering
	\includegraphics[scale=0.625, clip, trim=0 0 25 0]{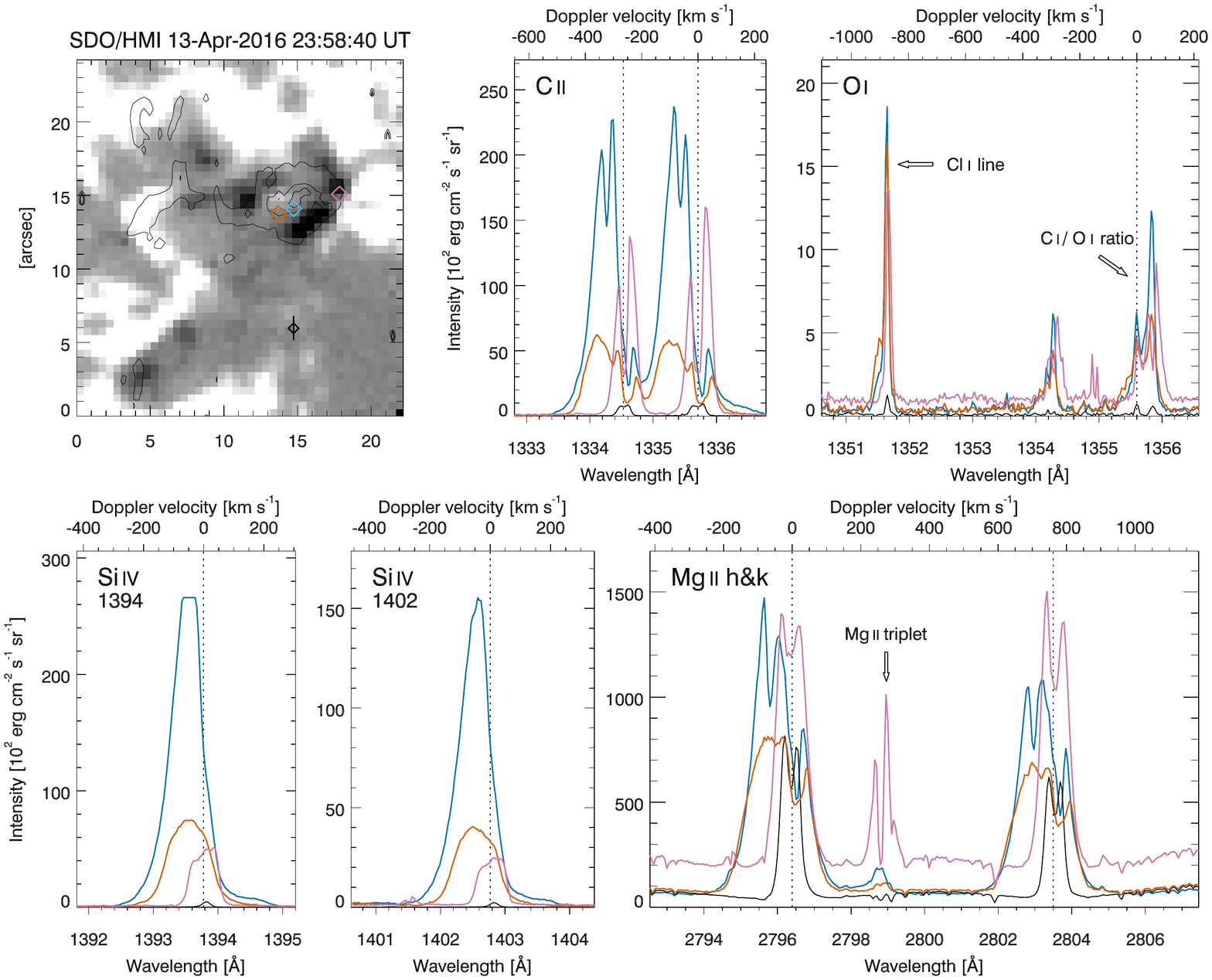}
	\caption{\textit{Top-left panel}: SDO/HMI LOS magnetogram, with overplotted contours of the \ion{O}{1} 1355.6~\AA{} line radiance. Some pixels, whose spectra are analyzed in detail, are indicated with colored diamonds. 
	The black diamond indicates the center of the quiet Sun area, used as a reference. The black segment indicates the six pixels along which the quiet Sun spectrum has been averaged. \textit{Other panels}: UV spectra from \textit{IRIS} observations. Each colored line refers to the spectrum of one of the colored pixels indicated in the top-left map. The black line is the reference spectrum in the quiet region. \label{fig_irisprofile}}
\end{figure*}

Elongated dark features departing from the EUV brightening site are also seen in the simultaneous images taken through the SDO/AIA 335~\AA{} filter (Figure~\ref{fig_jets}, third row). These 
structures, clearly visible in the online movie with the highest cadence of the SDO/AIA data, appeared to be cospatial to the surge-like ejection observed at 2796~\AA{}. See, for instance, the feature at X=$[10\arcsec, 18\arcsec]$, Y=$[10\arcsec, 20\arcsec]$ at 00:43:44~UT. The SDO/AIA 131~\AA{} channel (Figure~\ref{fig_jets}, fourth row) exhibits similar structures, although the contrast is not as clear as in the other wavelengths. However, the running difference images (see Figure~\ref{fig_jets}, fifth row), which have been obtained by subtracting each image with one 60~s before, allow identifying these structures as moving bright features. In this respect, the difference images from 00:43:44~UT to 00:47:55~UT make visible the evolution of a structure that expanded to the same extent as the surge observed in the 2796~\AA{} passband. At 00:43:44~UT, the ends of the moving bright feature were located at X=$15\arcsec$, Y=$10\arcsec$ and X=$25\arcsec$, Y=$20\arcsec$. They reached X=$20\arcsec$, Y=$35\arcsec$ and X=$30\arcsec$, Y=$45\arcsec$ at 00:47:55~UT. Such a motion is evident in the online movie. Moreover, a \textsc{Y}-shaped structure, cospatial with the jet-like feature, is recognizable in the SDO/AIA 131~\AA{} map at 00:52:05~UT and in the corresponding difference image. Such a feature is even more recognizable in the edge-enhanced image of the frame relevant to 00:51:51~UT in the online movie, which has been obtained using a difference of Gaussians filter to enhance the edges of the structure. Therefore, this jet-like feature underwent the same evolution as the surge-like ejection found in \textit{IRIS} 2976~\AA{} SJIs. 

On the other hand, a different kind of intensity enhancements took place in the region immediately next to the EUV compact brightenings. They appear as dimmings/brightenings in the difference images. In particular, some of them have an elongated configuration: see, i.e., the difference image at 00:41:38~UT in Figure~\ref{fig_jets} at X=$15\arcsec$, Y=$[13\arcsec, 22\arcsec]$, which shows one of them on the left edge of the EUV burst. Another example is provided in the SDO/AIA 335~\AA{} and 131~\AA{} filtergrams relevant to 23:58~UT in Figure~\ref{fig_synoptic} at X=$15\arcsec$, Y=$[15\arcsec, 20\arcsec]$, which reveal an elongated brightening to the right edge of the EUV burst. 

\subsection{Analysis of UV lines from \textit{IRIS} observations}

Figure~\ref{fig_irisbrightest} show the spectral profile of the \ion{Si}{4} 1402~\AA{} line in the brightest spatial pixel of each \textit{IRIS} raster scan. The total intensity of each spatial pixel has been calculated by integrating the intensity taken at each spectral sampling point in the range $1399~\mathrm{\AA} - 1406~\mathrm{\AA}$. We chose to show spectral profiles at the brightest pixels in each scan because the spatial evolution of the event occurs on a timescale that is much faster than the cadence of the \textit{IRIS} raster scans. It is not possible, hence, to see how spectral profiles of a particular location of the event change with time.

The first scan had an exposure time of 30~s, different from that used in the others (18~s), and the intensity reached the saturation level. In order to compare the profiles, the spectra have been calibrated into physical units ($\mathrm{erg}\:\mathrm{cm}^{-1}\:\mathrm{s}^{-1}\:\mathrm{sr}^{-1}$ per pixel) by using the \textsc{iris\textunderscore{}calib\textunderscore{}spec} routine. Thus, one can see that the maximum enhancement occurred during the third scan, when the brightest point had a full width at half maximum (FWHM) of $\sim 235 \,\mathrm{km\,s}^{-1}$. At that time, and also in the fourth scan, different plasma components seemed to be present in the resolution elements, as indicated by the presence of multiple peaks. The intensity and the FWHM had a peak between the third and fourth scans. Furthermore, the line exhibited red asymmetry, except during the second scan.

Figure~\ref{fig_irisprofile} displays the UV spectra acquired by \textit{IRIS} for some positions along the third raster scan, during the maximum emission enhancement. The spatial locations are chosen rather randomly across the UV burst to highlight different behaviour of the spectral profiles in the feature. They are taken in the central region of the burst and near the interface region between the opposite polarities down in the photosphere. In the plots, each colored line refers to the spectrum of the colored pixels with the same color, indicated in the top-left map of the same Figure. For comparison, we also present a reference average spectrum for six pixels along the \textit{y} direction in a quiet Sun area (black curve). 

The orange spectrum is taken in a pixel that exhibited a blueshift of about $100 \,\mathrm{km\,s}^{-1}$ in the \ion{C}{2}, \ion{Si}{4}, and \ion{Mg}{2} h\&k lines. Faint emission was found at the location of the \ion{Mg}{2} 2798.8~\AA{} triplet. The blue spectrum indicates that in the pixel there were blueshifts of the same amount as for the orange pixel in those lines, albeit with an increase of radiation of about a factor of 4 with respect to the latter. Also the \ion{Mg}{2} 2798.8~\AA{} triplet was stronger at this location. Furthermore, both \ion{C}{2} and \ion{Mg}{2} h\&k lines presented a double-peak configuration, with two peaks at $-80 \,\mathrm{km\,s}^{-1}$ and $-40 \,\mathrm{km\,s}^{-1}$. The \ion{Si}{4} lines, instead, were quite asymmetric. In the magenta spectrum, both \ion{C}{2} and \ion{Mg}{2} h\&k lines appeared to be at rest. A slight redshift is found in the \ion{Si}{4} lines, with a rather asymmetric profile. In addition, at this position the \ion{Mg}{2} triplet presented a strong enhancement, being the emission comparable to that of the \ion{Mg}{2} h\&k lines. 

The \ion{O}{4} 1401.2~\AA{} emission line is very weak next to the \ion{Si}{4} 1402~\AA{} line (Figure~\ref{fig_irisprofile}), and can not be reliably measured. We estimated an upper limit of 175 for the \ion{Si}{4} 1402~\AA{} / \ion{O}{4} 1401.2~\AA{} line ratio in the blue spectrum shown in Figure~\ref{fig_irisprofile}. If the large ratio is assumed to be due to a high density, this corresponds to densities of $\approx 10^{19} \,\mathrm{m}^{-3}$ applying the density-ratio curve of \citet{Peter:14}, after multiplying the ratio by a factor two as the curve is for the \ion{Si}{4} 1394~\AA{} / \ion{O}{4} 1401.2~\AA{} ratio, under the assumption that plasma is optically thin. This density is of the same order of the \textit{IRIS} bombs reported by \citet{Peter:14}. Unlike these events, however, the \ion{Ni}{2} 1393.3~\AA{} absorption feature is not present in the broad \ion{Si}{4} 1394~\AA{} profiles (Figure~\ref{fig_irisprofile}), suggesting there is no overlying cool material. 

\begin{figure}[t]
	\centering
	\includegraphics[scale=0.45, clip, trim=125 80 185 120]{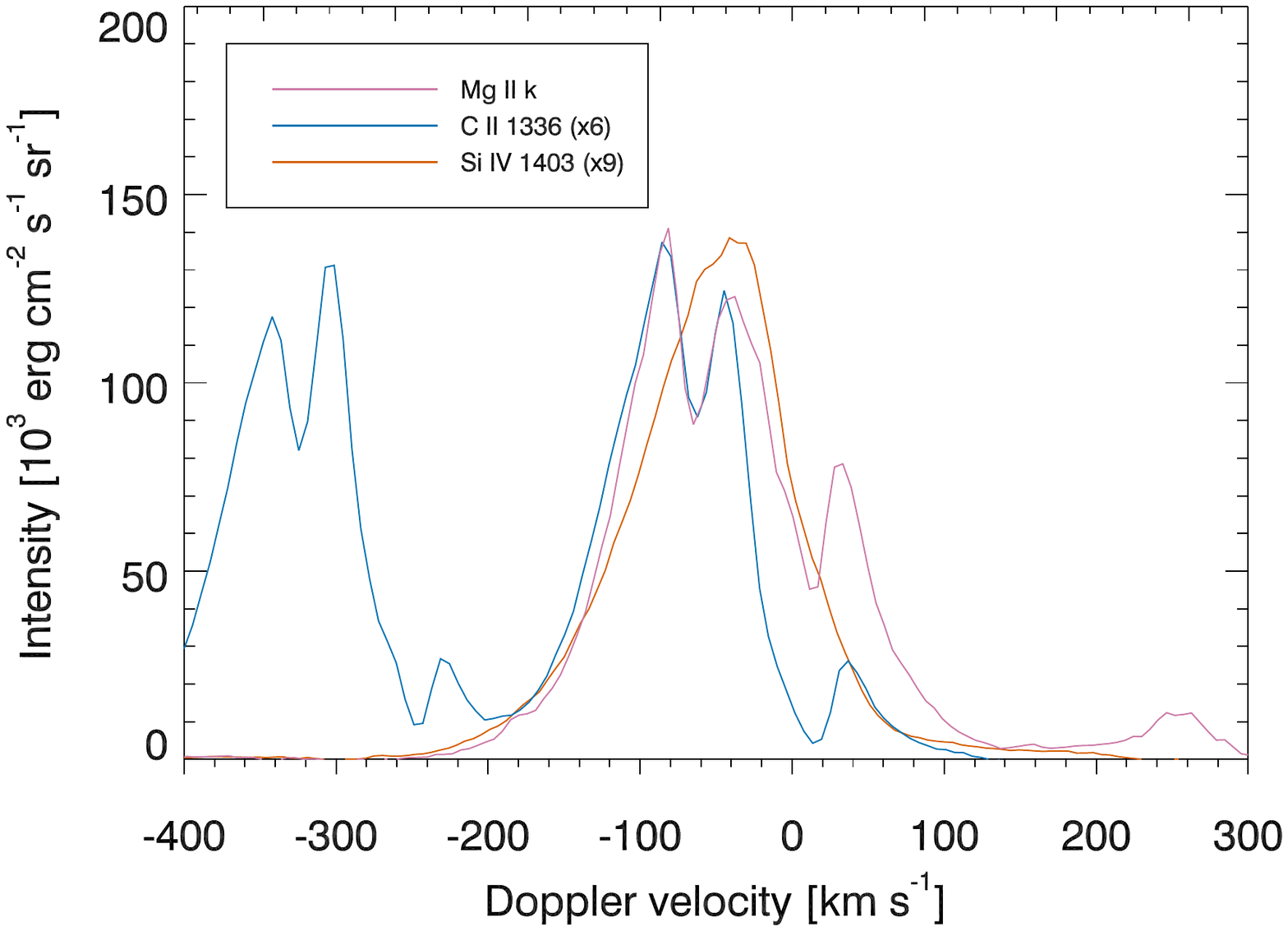}
	\caption{Graph showing the overplotted spectrum of the \ion{Si}{4} 1402, \ion{C}{2} 1335.7~\AA{}, and \ion{Mg}{2} h lines, relevant to the blue pixel in Figure~\ref{fig_irisprofile}. \label{fig_irissuper}}
\end{figure}

In order to understand if the UV lines were being formed in the same plasma, we have over-plotted in Figure~\ref{fig_irissuper} the line profile of \ion{Si}{4} 1402~\AA{}, \ion{C}{2} 1335.7~\AA{}, and \ion{Mg}{2} h for the pixel indicated with blue color in Figure~\ref{fig_irisprofile}. We have removed the continuum level from each profile to compare in a easier way the line wings. Figure~\ref{fig_irissuper} reveals that \ion{C}{2} 1335.7~\AA{} and \ion{Mg}{2} h line profiles had the same basic shape during the bursts, with those two peaks at $-80 \,\mathrm{km\,s}^{-1}$ and $-40 \,\mathrm{km\,s}^{-1}$, and a smaller peak at $+20 \,\mathrm{km\,s}^{-1}$ being common features of both line profiles. Such peaks at the same position point to the presence of distinct velocity components. Note that they were not apparent in the \ion{Si}{4} line. The overall shape of the profiles, including the wings, are similar if we take account of the self-reversals. Thus, the line profile comparison indicates that the lines are formed in the same plasma, but there are differences in the fine details of the line profiles, which may be related to small-scale dynamics.

Focusing again on Figure~\ref{fig_irisprofile}, in the spectral range of faint lines around \ion{O}{1} 1355.6~\AA{} we can notice a peculiar behaviour of the ratio between the intensity of the \ion{O}{1} 1355.6~\AA{} and \ion{C}{1} 1355.8~\AA{} lines. In the orange pixel, they have similar intensities. Surprisingly, in the blue and magenta pixels the \ion{C}{1} intensity is larger than in the \ion{O}{1} line, in particular in the blue spectrum it is twice as large. In addition, the magenta spectrum exhibits a third peak, being redshifted with respect to the \ion{C}{1} 1355.8~\AA{} line. The presence of a redshifted component in the magenta spectrum was also visible in the \ion{C}{1} 1354.2~\AA{} line. These peaks might also be due to a self-reversal in the \ion{C}{1} lines.

Moreover, in this same spectral window around \ion{O}{1} 1355.6~\AA{} line, the \ion{Cl}{1} 1351.6~\AA{}line, which is formed via a fluorescence effect driven by the \ion{C}{2} 1335~\AA{} line \citep{Shine:83}, shows a small hump to the blue wing of the line in the blue and orange pixel, the latter being more pronounced. 

With regard to the coronal lines, there was a very faint emission of \ion{Fe}{12} 1349.4~\AA{} line, which is compatible with a marginal detection characterized by a line width of the same order of the pure thermal FWHM (0.1~\AA{}). The other coronal line observed by \textit{IRIS} is the hot \ion{Fe}{21} 1354.08~\AA{}, which is generally only seen in flares and has a large thermal FWHM of about 0.41~\AA{}. It lies close to \ion{C}{1} 1354.29~\AA{}, which normally can be clearly distinguished due to its narrow width. However, in the burst spectra the \ion{C}{1} line is often broadened and can display extended wings, in particular to the blue. These features could be due to \ion{Fe}{21} emission, but comparisons with the nearby \ion{C}{1} 1355.8~\AA{} line demonstrate that the 1354.29~\AA{} line is rather consistent with only \ion{C}{1} emission, and hence we find no clear evidence for \ion{Fe}{21} emission from the burst. 

A more detailed analysis of the UV line properties observed during the evolution of the burst and in the other structures, such as the AFS and the surge-like ejections, will be provided in a follow-up paper.

\section{Discussion}

It is widely accepted that the emergence of magnetic flux is due to the buoyant rise of an $\Omega$-loop from a toroidal magnetic flux rope, embedded in the convective zone \citep{Parker:55,Fan:09}. While the magnetic field is reaching the photosphere, it suffers a distortion owing to the convective flows, which depends on the original field strength and twist of the structure \citep[see, e.g.,][]{Cheung:07,Sykora:15}. This causes an undulation of the rising flux tube that breaks in small bundles, leading to the serpentine appearance of the emerging field \citep{Cheung:10,Birch:16}. Elongated granules are observed in correspondence of the horizontal emerging fields, with dark alignments in between the opposite polarities, both in numerical models \citep{Cheung:07,Sykora:08,Tortosa:09} and high-resolution observations \citep{Guglielmino:10,Rolf:10,Ortiz:14,Centeno:17}. Indeed, the photospheric signatures observed during the evolution of the EFR that have been presented in this work fit very well with this scenario. We found dark, elongated features (dark alignments, following \citealp{Zwaan:08}) in between the emerging polarities of the EFR (see Figure~\ref{fig_hmi}). These structures are known to be present in the central part of an EFR, where the granulation looks fuzzy with the presence of such alignments. The dark alignments are thought to be caused by tops of magnetic loops passing through the photosphere, tracing the horizontal emerging fields of the EFR in the emergence zone \citep{Lites:98}. Such dark alignments are also clearly visible in the SOT/SP scan (Figure~\ref{fig_sot}) as well as in the \textit{IRIS} 2832~\AA{} continuum-like maps (Figure~\ref{fig_synoptic}, second column). The emergence zone was characterized by horizontal fields appearing as fragmented bundles (see Figure~\ref{fig_sot}).

Furthermore, it is worth stressing that SDO/HMI probably witnessed a flux cancellation event in the region of interest. In fact, a pore with positive polarity (P$^{+}$) disappeared. Then, it was replaced by a new pore with negative polarity (P$^{-}$), owing to the piling up of the emerging negative field of the EFR. This phenomenon is distinctly visible in the online movie relevant to Figure~\ref{fig_hmi}. This is a highly suggestive indication that a flux cancellation episode took place, resulting from the interaction and reconnection between the pre-existing field of the following polarity of the AR and the newly emerging EFR, rather than being due to flux retraction or decay.

With regard to the response of the upper atmospheric layers to flux emergence, we observed strong brightenings cospatial with the EFR in the chromosphere and in the TR, as well as in the corona, throughout the \textit{IRIS} observing sequence. The emission in the chromosphere shows a smooth enhancement during the flux emergence (see the lightcurves in the \textit{IRIS} 2796~\AA{} passband and in the SDO/AIA 1700 and 1600~\AA{} filters), whereas the energy release appears to be episodic in the TR and corona, with a duration of 5-10 minutes per burst (see the lightcurves in the \textit{IRIS} 1400~\AA{} passband and in the SDO/AIA EUV channels). At the same time, plasma surge-like ejections occur in all these atmospheric layers in the area of the EFR, and bi-directional plasma flows are found in UV lines at the same locations, with velocity up to $100 \,\mathrm{km\,s}^{-1}$. 

These signatures provide a rather compelling evidence that the event here analyzed is the result of magnetic reconnection between the emerging field and the pre-existing flux, in agreement with previous observations \citep[e.g.,][]{Guglielmino:12,Ortiz:16} and MHD numerical simulations \citep[e.g.,][and references therein]{Nobrega:16}. The reconnection releases bursts of energy in the chromosphere and TR, but also at coronal levels. As a matter of fact, \citet{Nobrega:17} developed a numerical model of magnetic flux emergence in a pre-existing magnetized medium to study the formation of surges and UV bursts. They showed that a non-stationary magnetic reconnection process gives rise to elongated thread-like structures, resembling crests that are ejected as plasmoids from the dome of the EFR, slightly away from the reconnection site. These surges follow the field lines of the ambient field, as already highlighted by other numerical simulations of surges \citep{David:15}. Closer to the reconnection point, almost at the edges of the burst, a hot \textsc{Y}-shaped jet occurs, with temperature of $T \sim 2-3$~MK, extending up to $\sim 20$~Mm, which can be hence observed in the corona. This could justify the observed elongated brightenings seen in the SDO/AIA EUV channels at a location immediately next to the burst we analyzed. 

The characteristics of the observed UV profiles also provide an indication for the occurrence of the plasmoid instability during the small-scale magnetic reconnection process. Indeed, the broad, multi-component UV profiles in the burst site might suggest the presence of several small plasmoids moving at different velocities \citep[see, e.g.,][]{Innes:15}. Moreover, \citet{Ni:15,Ni:16} showed that temperatures that are attained in these events ($\gtrsim 8 \times 10^4$~K) can be formed in a magnetic reconnection process in the low solar atmosphere, in which the slow- and fast-mode shocks at the edges of the plasmoids inside the current sheet regions are the main mechanism to heat the plasma and trigger the high temperature explosions. A similar conclusion has been recently reached by \citet{Luc:17}.

In this context, the compact brightness enhancement we observed is comparable to \textit{IRIS} bombs found by \citet{Peter:14} in terms of intensity, line width, and weakness of the \ion{O}{4}, although there is not chromospheric absorption in the \ion{Ni}{2} 1393.3~\AA{} line. The difference between those episodes and the brightening found in our observations is likely related with the larger height of the present burst in the solar atmosphere.

\begin{figure}[b]
	\centering
	\includegraphics[scale=0.45, clip, trim=80 20 240 40]{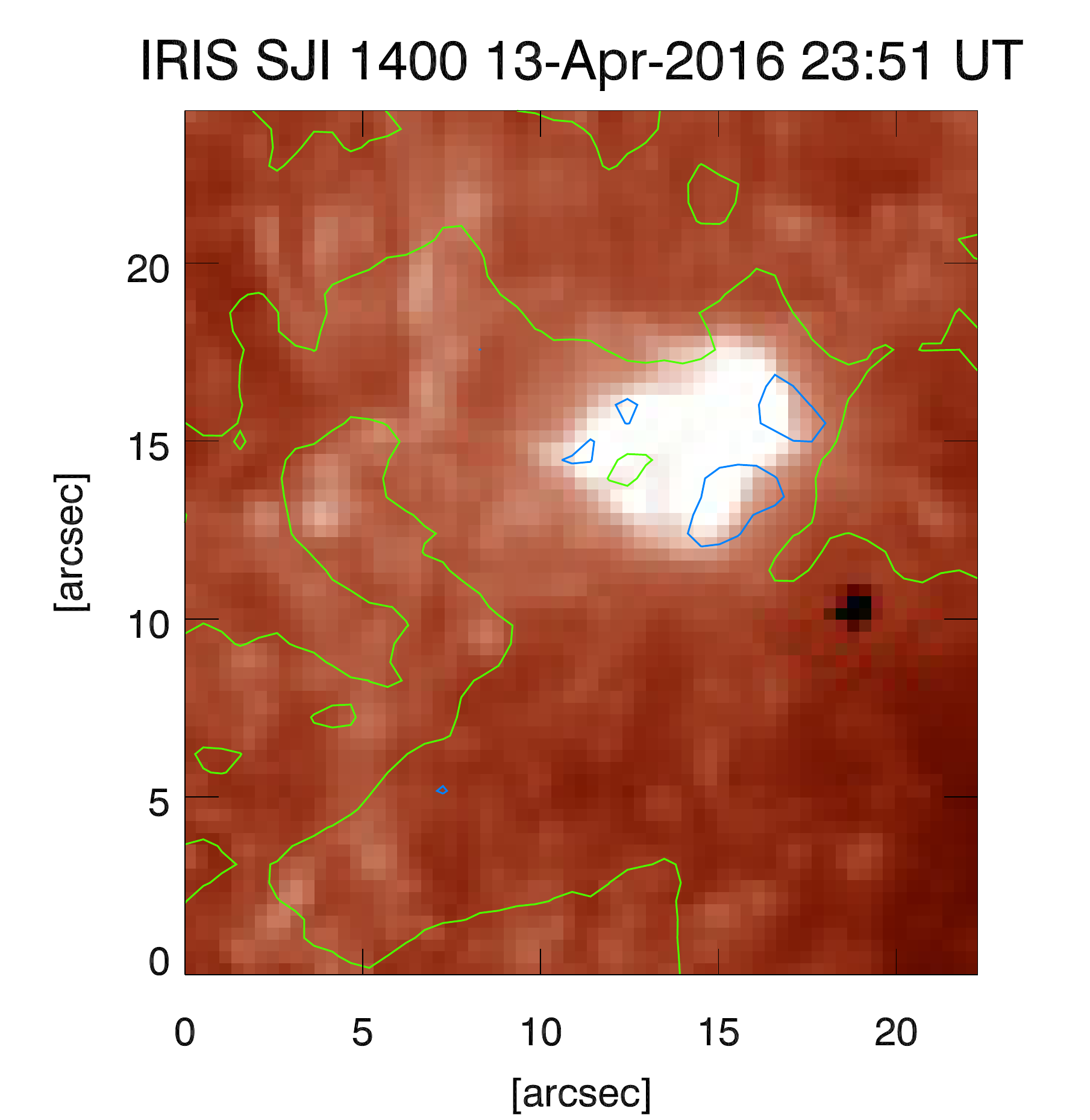}
	\caption{Snapshot of a sequence of \textit{IRIS} 1400~\AA{} SJIs, displaying the subFoV shown in Figure~\ref{fig_synoptic}, with overplotted contours of the SDO/HMI magnetograms closest in time (green/blue: +/- 100~G).\\
	(An animation of this figure is available in the online Journal.) \label{fig_irishmi}}
\end{figure}

Indeed, the burst here studied is not coincident with the polarity inversion line between the negative emerging field and the positive pre-existing flux in the cancellation site, but further to the East. This occurs throughout the \textit{IRIS} observations, as seen in the snapshot of the movie shown in Figure~\ref{fig_irishmi}, which displays a sequence of the \textit{IRIS} 1400~\AA{} SJIs with overplotted contours of the SDO/HMI magnetograms closest in time. Actually, MHD simulations indicate that the reconnection site is higher in the dome formed by the magnetic domain of the EFR interacting with the ambient field, so that it is displaced with respect to the photospheric site where the colliding opposite polarities cancel each other \citep{Nobrega:17}. Furthermore, a higher location of the reconnection site in the solar atmosphere with respect to the event studied by \citet{Peter:14} might explain the lack of chromospheric absorption. 

However, note that the emergence events modelled by \citet{Nobrega:17} has a flux content of about a factor of 20 less than the EFR we observe. Also, the effects of numerical viscosity may play a role in smoothing some physical effects. These facts, together with the stronger ambient magnetic field in which the EFR is embedded, may explain the longer duration of the burst we observed, which consists of repeated reconnection episodes during $\sim 3$~hrs, compared to the simulated one, which has a lifetime of about 10~minutes.  

Other information can be inferred from some UV signatures, as briefly analyzed in this work. In our observations, we found \ion{Mg}{2} triplet emission at the burst location. The emission in the \ion{Mg}{2} triplet is rare, caused by a steep temperature increase in the lower chromosphere at high electron density ($> 10^{17}\,\mathrm{m}^{-3}$), as demonstrated by \citet{Pereira:15}. They also explained that the \ion{Mg}{2} triplet emission occurs predominantly in the wings if the heating occurs deeper down and is covered by cooler material, whereas the emission takes place in the line core when the heating occurs higher in the column range. The observed \ion{Mg}{2} triplet emission is located in the wings at the periphery of the UV burst, similar to that observed in EBs \citep{Hansteen:17}, whereas it is situated mainly in the line core at the burst center. This further supports that the UV burst studied here occurs slightly higher in the atmosphere than \textit{IRIS} bombs, seemingly at chromospheric heights. Moreover, it provides an additional explanation to the lack of chromospheric absorption lines in the \ion{Si}{4} profiles.

Note that the evolution of the UV burst shows many extending dynamic loop-like features that are rather similar to those seen by \citet{Huang:15} in the footpoints of cool TR loops rooted in mixed-polarity flux regions. Nevertheless, the observed spectra do not show clear EE-type \ion{Si}{4} 1402~\AA{} line profiles with enhanced wings like those found by \citet{Huang:15}, suggesting that this UV burst might be occurring at temperatures lower than \ion{Si}{4} temperature. In this assumption, the \ion{Si}{4} spectra do not react on the bulk velocity of plasma but only show result of heating therein. However, the EEs studied by \citet{Huang:15} seem to be related to magnetic flux cancellation between two pre-existing loop systems, rather than to the interaction between ambient fields and newly emerging flux. This point could reflect the partly different spectra observed in the TR in these dynamic events.

Finally, particularly intriguing is the behaviour of the ratio between the \ion{C}{1} 1355.8~\AA{} and \ion{O}{1} 1355.6~\AA{}. We found that this ratio increased in the location of the UV burst, with values up to $\sim 2$. \citet{Cheng:80} discovered that the \ion{C}{1}/\ion{O}{1} ratio was remarkably enhanced during flares, compared to the typical values of $0.5 - 1$ in the quiet Sun. \citet{Cheng:80} suggested that this variation might be due to an electron density enhancement by a factor of $\sim 50$ of the chromospheric plasma. Presently, a clear explanation of the physical implications of the changes in the \ion{C}{1}/\ion{O}{1} ratio is lacking \citep{Lin:15}, but we note that UV bursts have been demonstrated to have high electron densities at TR temperatures \citep{Doschek:16}. Further investigations are required to clarify this issue.

\section{Conclusions}

We have studied the relationship between an EFR observed in the photosphere, embedded in a unipolar field, and burst-like events with signatures in the UV and EUV. We have provided evidence that intensity enhancements and plasma ejections occur as a result of magnetic reconnection, which leads to a simultaneous flux cancellation at the photospheric level. These findings support the link between UV bursts and surges, when the flux amount involved in the cancellation events is large enough \citep{Shelton:15}. 

In this respect, it is worthwhile to note that this UV burst is long-lived and spatially complex, different from most bursts which are simple compact bright points, arising when a single, small magnetic flux concentration cancels against stronger field. This is likely due to the fact that, in the present case, the cancelling magnetic features are larger, and also that multiple fragments cancel over time.  

The presence of an AFS reflects the serpentine geometry of the emerging flux. This provides a further suggestion that the long-lived dynamic behavior of the UV burst, which is well represented by the pulses in the light curves, is actually resulting from magnetic reconnection between emerging and pre-existing flux systems. In fact, it may occur between the threads of an AFS and pre-existed coronal loops when some mechanisms force them to encounter each other and provide inflows of the reconnection \citep[see, e.g.,][]{Huang:18}. The flows along the reconnected loops appear to have various speeds of about $40 - 80 \,\mathrm{km \,s}^{-1}$, possibly due to intermittent reconnection, similarly as in the UV burst here analyzed. In this regard, the long-lived brightening might be representative of the response of the pre-existed flux in the different atmospheric layers, while its impulsive behavior might be ascribed to the reconnection between the emerging and pre-existing field lines.

Another key aspect of our observations is the very strong SDO/AIA coronal signature of the UV burst. The hot explosions reported by \citet{Peter:14} do not have counterpart in the SDO/AIA EUV channels, while the EBs analyzed in AFSs by \citet{Vissers:15a} and the recurrent EEs studied by \citet{Gupta:15} show only brief signatures in the 171 \AA{} and 193 \AA{} channels. In contrast, the burst investigated here does reveal intensity enhancements in all of the atmospheric levels.  

A comparison with numerical simulations indicate that the magnetic topology overlying UV bursts determines its influence at coronal levels. We do see small-scale flux cancellation in the photosphere, but observational evidence that the reconnection site is higher in the atmosphere than usually observed in \textit{IRIS} bombs and other small-scale energy release episodes, provides an explanation for the presence of coronal intensity enhancements and jet-like ejections. This suggests that, if triggered in a favourable magnetic configuration, UV bursts may play a role in the coronal heating as well, as pointed out by \citet{Chitta:17}.

In the follow-up paper, we will analyze the UV line profiles during the six \textit{IRIS} scans, performing a statistical analysis of the profiles relevant to the burst. Moreover, we will investigate the dynamics related to the AFS as observed by \textit{IRIS}.   

We expect that progress in understanding the relationship between small-scale flux emergence episodes in the photosphere and energy release phenomena in the upper atmospheric layer will be achieved by benefitting from the high spatial resolution and continuous temporal coverage provided by the next generation of solar observatories, such as the Solar Orbiter space mission \citep{Muller:13} and the large-aperture ground-based telescopes DKIST \citep[Daniel K. Inouye Solar Telescope,][]{Keil:10} and EST \citep[European Solar Telescope,][]{Collados:10}.

\acknowledgments

The authors would like to thank the anonymous referee for his/her helpful comments. The research leading to these results has received funding from the European Commissions Seventh Framework Programme under the grant agreement no.~312495 (SOLARNET project) and from the European Union's Horizon 2020 research and innovation programme under grant agreement no.~739500 (PRE-EST project). This work was also supported by the Italian MIUR-PRIN grant 2012P2HRCR on \textit{The active Sun and its effects on Space and Earth climate}, by the Universit\`{a} degli Studi di Catania (PRIN MIUR 2012) on \textit{The active Sun and its effects on Space and Earth climate}, by the Instituto Nazionale di Astrofisica (PRIN INAF 2014), and by Space Weather Italian COmmunity (SWICO) Research Program. PRY acknowledges funding from NASA grant NNX15AF48G, and he thanks ISSI Bern for supporting the International Team Meeting ``Solar UV bursts -- a new insight to magnetic reconnection''. \textit{IRIS} is a NASA small explorer mission developed and operated by LMSAL with mission operations executed at NASA Ames Research center and major contributions to downlink communications funded by ESA and the Norwegian Space Centre. \textit{Hinode} is a Japanese mission developed and launched by ISAS/JAXA, with NAOJ as domestic partner and NASA and STFC (UK) as international partners. It is operated by these agencies in co-operation with ESA and Norwegian Space Centre. The SDO/HMI and SDO/AIA data used in this paper are courtesy of NASA/SDO and the HMI and AIA science teams. Use of NASA's Astrophysical Data System is gratefully acknowledged.

\facility{IRIS, Hinode (SOT), SDO (HMI, AIA)}

\end{document}